\documentclass[journal]{IEEEtran}
\ifCLASSINFOpdf
\else
\fi
\usepackage[cmex10]{amsmath}
\usepackage{booktabs}     

\usepackage{stfloats}
\usepackage{graphicx}
\usepackage{color}
\usepackage[dvipsnames]{xcolor}
\usepackage{placeins}
\usepackage{float}
\usepackage{tabularx,colortbl,multirow}

\usepackage{graphicx}
\usepackage{subfigure}
\usepackage{enumerate}
\usepackage{enumitem}
\usepackage{amsmath,amssymb,amsfonts,amsthm,tikz,caption}

\usepackage{mathtools}
\DeclarePairedDelimiter\abs{\lvert}{\rvert}

\usepackage{bm}

\renewcommand\eqref[1]{(\ref{#1})}

\usepackage{tikz}
	\usetikzlibrary{shapes,decorations,decorations.pathreplacing}
	\usetikzlibrary{arrows,shapes,positioning,calc,fadings,patterns}
	\usetikzlibrary{decorations,decorations.pathreplacing,decorations.pathmorphing}
\usepackage{tikz-3dplot}

\usepackage{booktabs}     
\usepackage{threeparttable}
\usepackage{amssymb}
\usepackage[style=ieee] {biblatex}
\bibliography{Biblio} 

\makeatletter
\newcommand{\vast}{\bBigg@{4}}
\newcommand{\Vast}{\bBigg@{5}}
\makeatother

\usepackage{siunitx}

\begin{document}
%
\title{  FDTD-Based Diffuse Scattering and Transmission Models for Ray-Tracing of Millimeter-Wave Communication Systems 
}

%

\author{Stefanos~Bakirtzis,~\IEEEmembership{Student~Member,~IEEE}, Takahiro~Hashimoto  and~Costas~D.~Sarris,~\IEEEmembership{Senior~Member,~IEEE}
\thanks{Manuscript received XX XX, 2020; revised XX XX, XXXX; accepted
XX XX, XXXX.}
\thanks{The authors are with the Edward S. Rogers Sr. Department of Electrical and Computer Engineering, University of Toronto, Toronto, ON M5S 3G4, Canada (e-mail: stefanos.bakirtzis@mail.utoronto.ca; costas.sarris@utoronto.ca).}
\thanks{Color versions of one or more of the figures in this paper are available
online at http://ieeexplore.ieee.org.}
\thanks{Digital Object Identifier: XXXX}
}

\maketitle


\begin{abstract}

At millimeter-wave frequencies, diffuse scattering from rough surfaces is an important propagation mechanism. Including this mechanism in  radio propagation modeling tools, such as ray-tracing, is a key step towards realizing accurate propagation models for 5G and beyond systems. We propose a two-stage solution to this problem. First, we model reflection and transmission  through rough slabs, such as doors, walls and windows with the FDTD method. Our results indicate the influence of roughness and whether this influence is measurable, either reducing the magnitude of the reflected and transmitted waves, or (most importantly) generating diffuse scattering components. In the latter case, the surface effectively acts as a secondary source, whose pattern is computed by full-wave analysis. Then, this pattern is embedded in a ray-tracer, enabling the computation and tracing of diffuse scattering field components. We demonstrate this approach in the ray-tracing analysis of a 28 GHz indoor environment. 

\end{abstract}


%
\IEEEpeerreviewmaketitle

\section{Introduction}
%
%
%
%

\IEEEPARstart{A}{t} millimeter-wave (mm-wave) frequencies, mm-size anomalies on the surfaces of indoor and outdoor environments are comparable to the wavelength. Therefore, their  influence on the propagation of mm-waves is much more pronounced than on microwaves in established wireless systems (cellular communications and Wi-Fi).

The conventional approach of accounting for the presence of roughness in radio propagation modeling is to modify the standard Fresnel reflection coefficient by an exponential factor \cite{janaswamy2001radiowave, landron1996comparison}. However, as the surface height variation due to roughness becomes comparable to or even larger than the wavelength, an incident plane wave produces scattered fields in multiple directions in addition to the specular. This is the effect of diffuse scattering, which has important implications in propagation studies for 5G systems.


From the perspective of wireless propagation modeling, it is essential to develop a tool that can accurately model diffuse scattering in general geometries, without any limiting assumptions, combined  with an efficient propagation method, such as ray-tracing (RT). Recent studies in THz communications focused on reflection from one boundary \cite{THz2019, THz}, using the Beckmann-Kirchhoff (B-K) \cite{Beckmann1987scattering} approximation to estimate diffuse scattering. However, this model is only valid under certain assumptions regarding surface characteristics (the correlation length of the surface must be much greater than the wavelength). More importantly, at those cases, the objects found in an indoor environment can be modeled as semi-infinite spaces due to the high losses inside material slabs. 
 
In the frequency bands used in 5G systems, such as the 28 GHz band, this assumption does not hold, as losses are considerably smaller. Therefore, reflecting surfaces, such as walls and doors, should be modeled as slabs formed by two rough interfaces \cite{langen1994reflection, ITU}, rather than semi-infinite spaces. In this case, the transmitted fields may also be significant and necessary to compute.

Full wave analysis provides an accurate way to evaluate diffuse reflection and transmission in non-trivial multi-layer cases, where the analytical treatment is not feasible. It can be applied to capture both the attenuation in the specular, as well as the entire scattering pattern of an arbitrarily complex geometry. To that end, we formulate a two-dimensional (2-D) FDTD model, based on the multilayered Total-Field/ Scattered-Field (TF/SF) approach proposed in \cite{smith2008total}, to simulate the  propagation of mm-waves through rough slabs. Instead of solving the single rough interface problem, we exploit the versatility  of FDTD  to identify additional diffuse paths that need to be further traced. Motivated by the concept of Bidirectional Scattering Distribution Functions (BSDF), utilized in computer graphics to describe diffuse scattering from surfaces \cite{CS-RT}, we use FDTD results to deduce compact models \cite{degli2007measurement} of the reflection and transmission coefficients from rough slabs. Finally, we integrate our FDTD model with an RT simulator, in order to enhance its accuracy in the challenging situation that involves strong diffuse scattering components.

The outline of this paper is as follows; first, in Section II, we provide a brief description of the topic of roughness in electromagnetics and we discuss some of the existing models and methods used to evaluate diffuse scattering. Next, in Section III, we describe the FDTD-based model we used to characterize reflection and transmission through rough slabs. In section IV, we provide our numerical results and in section V, we outline how these results can be used to fit diffuse scattering patterns, that can replace the Fresnel reflection and transmission coefficients of rough slabs \cite{degli2007measurement}. In section VI, we describe how we embedded the FDTD reflection and transmission coefficients in an RT simulator,  and  we provide results for an indoor propagation scenario at 28 GHz.

\begin{figure}[h]
\centering
    {\includegraphics[width =\columnwidth]{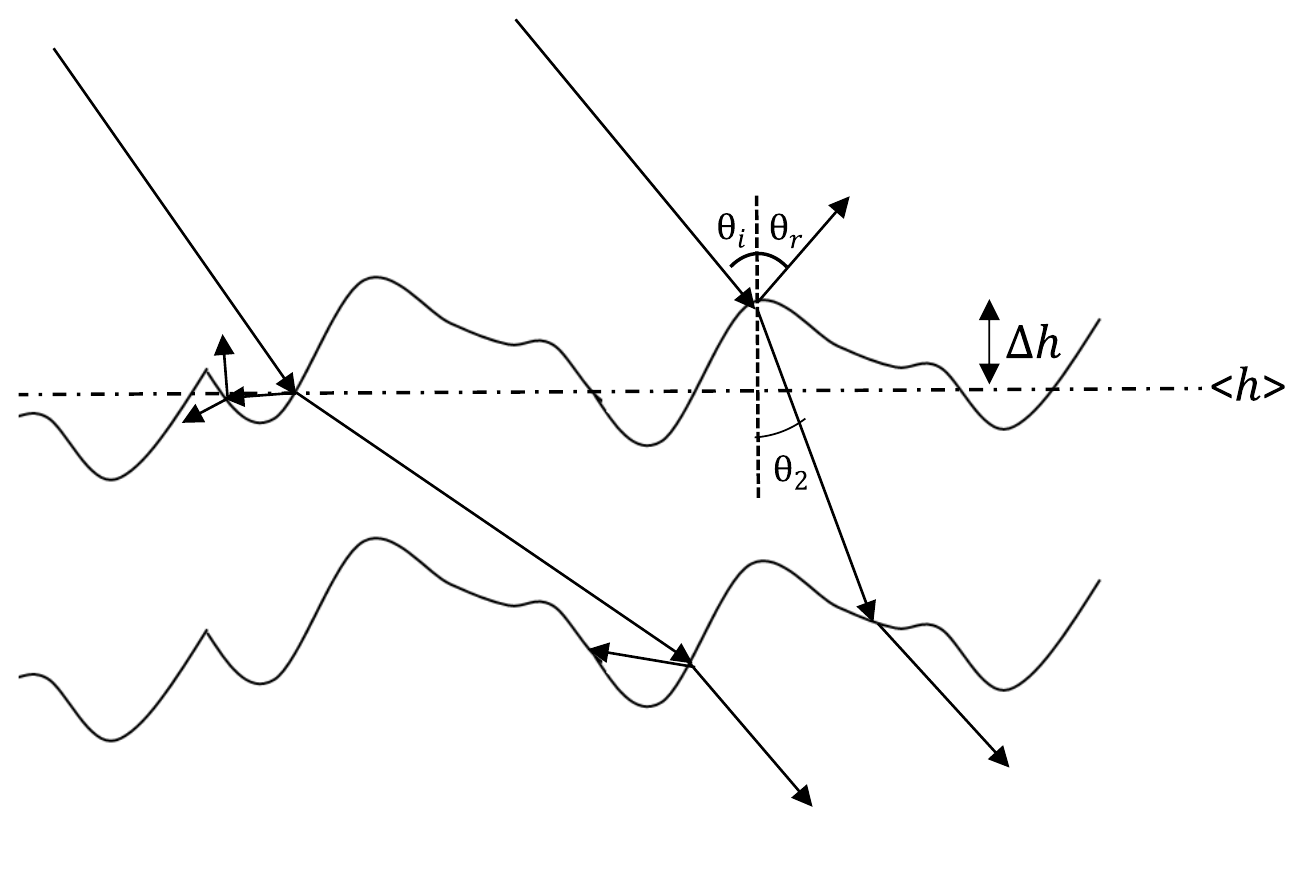}}
\caption{{Rays propagating through a rough slab. }}
\label{fig:Scattering}
\end{figure}

\section{Surface Roughness and Propagation Modeling}

 When a plane wave impinges onto a rough surface, different parts of a wavefront encounter the surface at a different height. Hence, each  scattered component has a phase difference, with respect to the phase of the reflected component from a flat surface at mean height $\langle h  \rangle $, given by \cite{Pinel}: 
 
 \begin{equation}
    \Delta \phi_R= 2k\Delta h\cos\theta_i
     \label{eq:Delatphi_R}
 \end{equation}
 where $k = 2 \pi / \lambda$ is the wavenumber, $ \theta_i$ is the  angle of incidence and $ \Delta h = h_1 - \langle h  \rangle $ is the height variation around the mean surface height $\langle h  \rangle $,  as shown in Fig.~\ref{fig:Scattering}. The total reflected field from the rough surface is the superposition of multiple scattered components, which can  interfere either constructively or destructively. The Rayleigh criterion, which defines the degree of  roughness, requires that for the interference to be significant, the standard deviation of the phase difference should be higher than $ \pi$/2. Assuming a zero mean height, the critical rms height in reflection, above which a random surface can be considered as  rough, is:
 \begin{equation}
     \sigma_{h_{c,R}}=  \frac{\lambda}{8\cos\theta_i}\
      \label{eq:Hc_R}
 \end{equation}

In a similar manner, one may obtain the phase for the transmitted components and the critical rms height for the transmission \cite{Pinel}:

 \begin{equation}
    \hspace{5mm} \Delta\phi_T =  k\Delta h(n_1\cos\theta_i - n_2\cos\theta_2)
      \label{eq:Delatphi_T} 
 \end{equation}

 \begin{equation}
     \sigma_{h_{c,T}} =   \frac{\lambda}{4(n_1\cos\theta_i - n_2\cos\theta_2)} 
     \label{eq:Hc_T}
      \end{equation}
\vspace{2mm}

\begin{figure}[h]
\centering
    {\includegraphics[scale = 0.65]{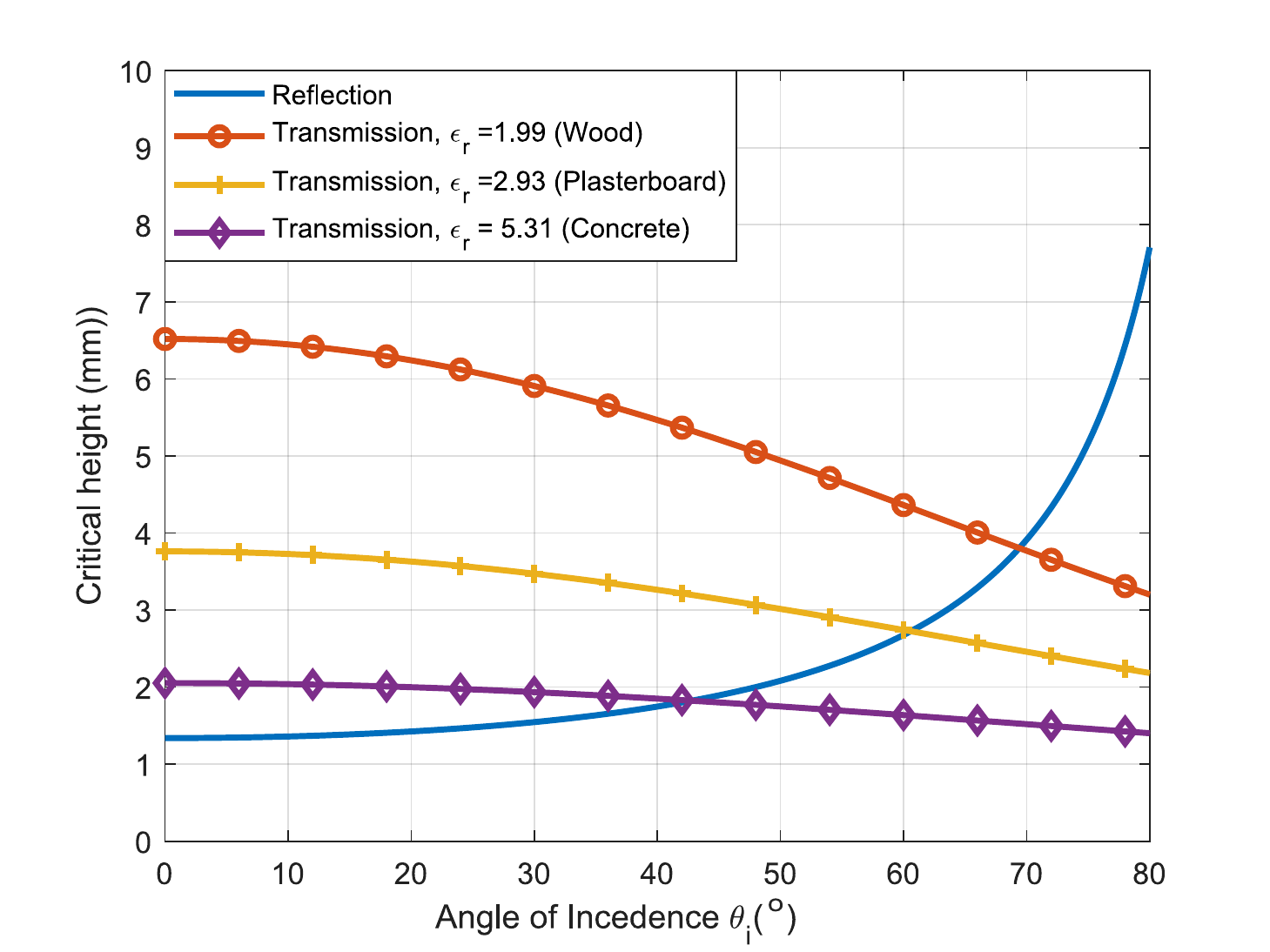}}
\caption{{Critical height in reflection and transmission for a wave propagating from air to a dielectric at 28 GHz. Three dielectric cases (wood, plasterboard, concrete) are presented.}}
\label{fig:Critical_Heights}
\end{figure}

From~(\ref{eq:Hc_R}) and ~(\ref{eq:Hc_T}), it is evident that the roughness perceived by the propagating wave depends on its frequency. Hence, for mm-waves, whose wavelength becomes comparable to small variations of the surface height, the role of roughness is substantial. Furthermore, the impact of roughness on reflection and transmission is different. In general, for materials with similar indices of refraction, the critical height is higher in transmission. In Fig~\ref{fig:Critical_Heights}, we show the critical height in reflection and transmission at different angles of incidence, for a plane wave propagating from air to three different dielectrics (wood, plasterboard, concrete) at 28 GHz. The critical height in reflection remains the same for all the materials, as it only depends on the wavelength and the angle of incidence. In transmission, the higher the relative permittivity, the smaller the critical height.

In remote sensing \cite{Ishimaru,Tsang, Bulks}, asymptotic models, such as the Kirchhoff Approximation (KA)  or the Small Perturbation Method (SPM), have been used to estimate the scattering fields from a rough surface. These methods provide an analytical solution to the scattering problem, under some simplifying assumptions \cite{Pinel}. Additionally, they  neglect phenomena, such as shadowing and multiple bouncing. Numerically exact methods have also been studied. In \cite{hastings1995monte}, a conformal FDTD method was used to compute the Radar Cross Section (RCS) of a single rough interface. In this paper, we build on this approach, solving the two-layer rough surface problem and evaluating both the reflection and the transmission for the two-layer structure. In particular, we solve the rough slab problem, as it is more important from the perspective of  communication channel modeling (e.g. walls, doors and windows). 

The problem of modeling diffuse scattering has also been studied in the context of wireless communications. For example, in \cite{degli2007measurement},  an effective roughness (ER) model was introduced to capture the diffuse scattering on non-uniform building walls at 900 MHz. However, the ER model does not correspond to actual, but to ``effective" roughness, caused by random variations of the  surface height, but to irregularities in buildings facets, such as windows and balconies. 

\section{Problem Formulation}

\subsection{Rough surface generation}

We follow the method introduced in \cite{Thorsos}, to generate zero mean, Gaussian random surfaces. The generated zero mean Gaussian surfaces are characterized by the standard deviation of their height, which is also the  rms height  of the surface $\sigma_h$, and their correlation length, $l_c$ (horizontal distance at which the correlation between the height at two points on the surface drops to $1/e$). 

We consider surfaces of a total length $L = N\Delta x$, where $N$ is the number of sampling points on the rough surface, and $\Delta x $ is the spacing between two consecutive points. For convenience, the value of $\Delta x $ is chosen to be equal to the cell size of FDTD along the surface. 
 
To generate a random surface, we associate each FDTD cell with a random sample from the normal distribution $N(0,1)$, giving rise to a Gaussian rough surface. Then, we take the inverse Fourier Transform of the spectrum to determine the height profile:

 \begin{equation}
      f(x_n) = \frac{1}{L} \sum_{m= -\frac{N}{2}} ^{\frac{N}{2}} F(K_m)\exp(-jK_m x_n)
      \label{eq:Height_prof}
 \end{equation}

 \noindent where  $K_m= 2 \pi m/ L$ and $x_n = n\Delta x$ ($n = 1..N)$ are the sampling points on the rough surface. The Gaussian spectrum $W(K_m)$ and the Fourier coefficients $ F(K_m)$ are defined as \cite{Thorsos}:

 \begin{equation}
    \hspace{-38 mm}  W(K_m) = \frac{ \sigma_h^2l_c }{2\sqrt\pi} \exp( - \frac{ K_m^2l_c^2}{4}) 
      \label{eq:Spec}
 \end{equation}

\begin{multline}
      F(K_m)   = \sqrt{2\pi L W(K_m)}  \\
   \cdot \vast\{
  \begin{aligned}
    &\frac{1}{\sqrt2}[N(0,1) -j N(0,1)] && \text{if $-\frac{N}{2}\leq m \leq \frac{N}{2}$  }\\
    &N(0,1) && \text{if $m = 0$, $\frac{N}{2}$} \\
  \end{aligned} 
  \label{eq:Fourrier}
\end{multline}

\subsection{FDTD-based modeling of  reflection and transmission from multi-layer rough surfaces}

We employ the multilayered TF/SF method proposed in \cite{smith2008total},  to generate an obliquely incident plane wave impinging onto a rough slab. In what follows, we present the analysis for the TE case. By duality, similar analysis can be performed for the TM case.

\begin{figure}[h!]
\centering
    {\includegraphics[scale=0.58]{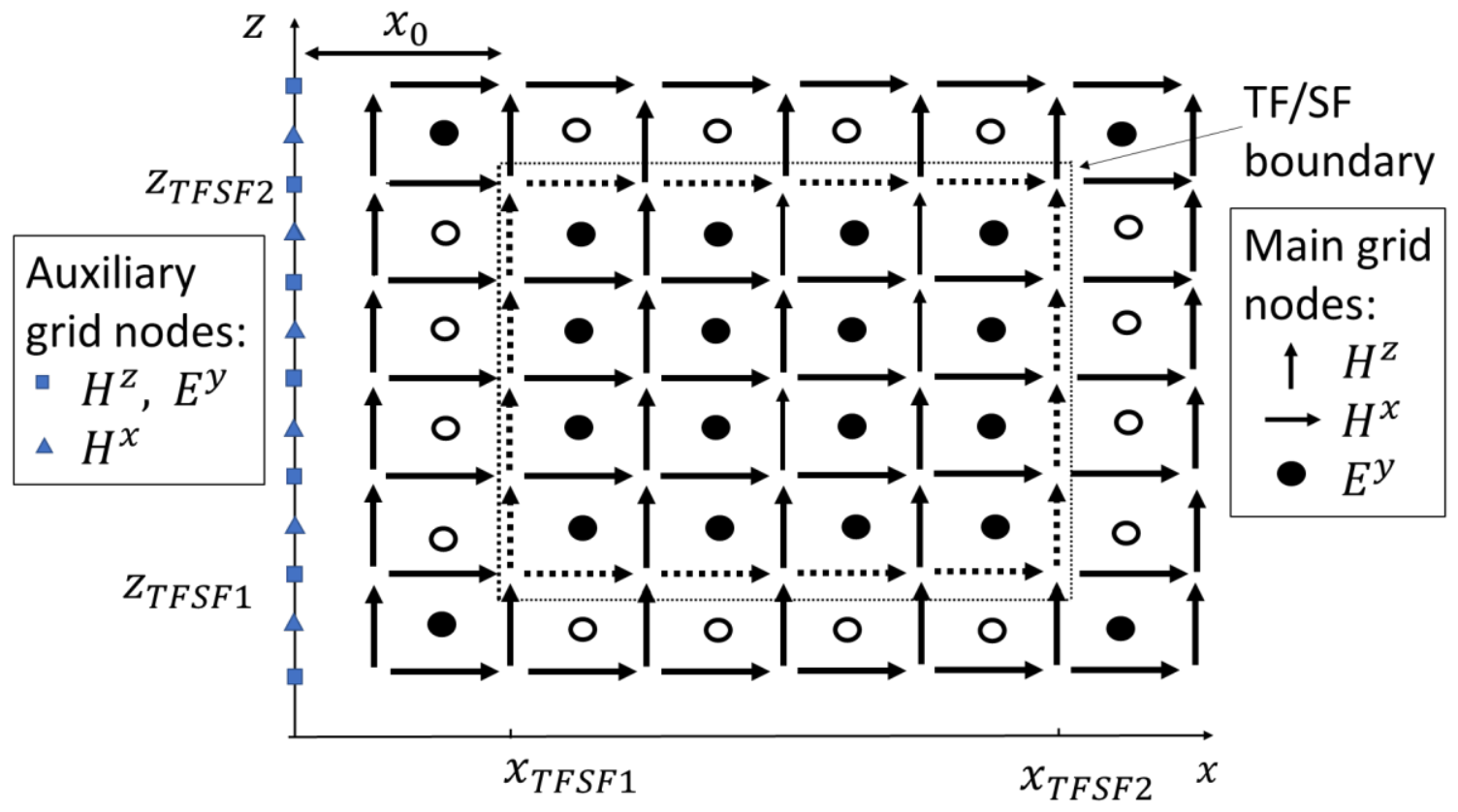}}
\caption{{Auxiliary and main FDTD grid.}}
\label {fig:FDTD_grid}
\end{figure}

At $x=0$, parallel to the $z$-axis, we set up a one dimensional TEM auxiliary grid, which will provide the required correction terms for the TF/SF boundary nodes of the main grid  \cite{taflove}. At each interface, the auxiliary grid ``sees" the same reflection and transmission coefficients as an obliquely incident wave. As shown in Fig.~\ref{fig:FDTD_grid}, the nodes of the auxiliary grid are aligned with the nodes of the main grid, along the $z$-axis.


Before performing the simulation for the main grid, we execute the 1-D auxiliary grid simulation and store the values of every field component at each time step. These values will be used to obtain the proper correction terms for the TF/SF boundary nodes, shown in red in Fig.~\ref{fig:FDTD_grid},  at each spatial point and each time step. Due to the phase matching condition, the tangential phase velocity of the plane wave remains the same at every layer, equal to $u_{p_x}= c_0/\sin\theta_i$, where $c_o$ is the speed of light in vacuum. To compute a correction term  at a given time step for a node at the TF/SF boundary (dashed arrows and hollow dots), we introduce a time delay to the stored values of the corresponding node of the auxiliary grid. For a node located at $x_0$, the time delay to introduce is $ \Delta\tau= x_0 u_p/\sin\theta_i$. The update equations of the 1-{D} auxiliary grid, for the TE case and for lossy media, are as follows \cite{smith2008total}:

\begin{flalign}
    \hspace{-18mm} H_k^{x, n+\frac{1}{2}}   =     H_k^{x,n-\frac{1}{2}} - \frac{\Delta t}{\mu_0 \Delta z} ( E_{k+\frac{1}{2}}^{y,n} - E_{k-\frac{1}{2}}^{y,n} )
 \label{eq:Aux1}
\end{flalign}

\begin{multline}
    E_{k+\frac{1}{2}}^{y, n+1}   =   \frac{A_-}{A_+}  E_{k+\frac{1}{2}}^{y,n} \\+ \frac{\Delta t}{\epsilon_0(\epsilon_{r_{i},k+\frac{1}{2}}' - \sin^2\theta_i) \Delta z A_+} ( H_{k+1}^{x,n+\frac{1}{2}} - H_{k}^{x,n+\frac{1}{2}} )
 \label{eq:Aux2}
\end{multline}

\begin{figure}[t]
\centering
    \subfigure[Geometry of the problem assuming flat interfaces.]
    {\label{fig:Oblique_flat}
    \includegraphics[scale=0.99]{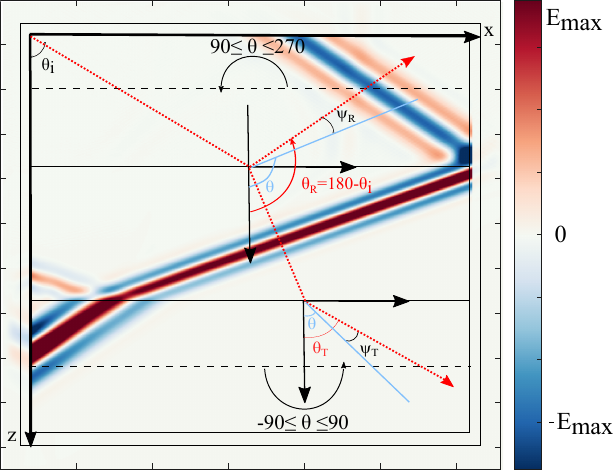}}
    \subfigure[Geometry of the problem assuming rough interfaces.]
    {\label{fig:Oblique_Rough}
    \includegraphics[scale=0.99]{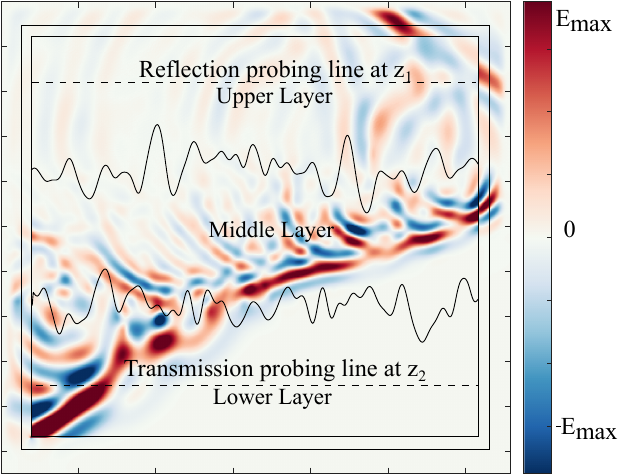}}
\caption{A two-layer TF/SF model, simulating the oblique incidence onto (a) a smooth and (b) a rough slab.}
\label{fig:Oblique}
\end{figure}

\vspace{2mm}

\begin{equation}
    \hspace{-33mm} H_{k+\frac{1}{2}}^{z, n+\frac{1}{2}}   =     H_{k+\frac{1}{2}}^{z,n-\frac{1}{2}} - \frac{2\sin\theta_i}{\eta_o}  E_{k+\frac{1}{2}}^{y,n} 
 \label{eq:Aux3}
\end{equation}

\noindent  where  $A_{-,+} = 1 \mp \frac{\sigma_{i, k+\frac{1}{2}} \Delta t}{2 \epsilon_o(\epsilon_{r_{i},k+\frac{1}{2}}' - \sin^2\theta_i)} $,   $ \sigma_i$ and $ \epsilon_{r_i}^{'} $ are the conductivity and the real part of the relative permittivity of the $i$-th layer, respectively, and $\eta_0$ is the free space impedance. In our case, for the uppermost and the lowermost layers (see Fig.~\ref{fig:Oblique_Rough}), we have $ \epsilon_{r_{1,3}}^{'} = 1$ and $\sigma_{1,3} = 0$. For the middle layer, the values of $\epsilon_{r_2}^{'}$ and $\sigma_2$  are set according to the type of the material examined.

\begin{figure}[h]
\centering
\hspace*{0 cm} 
    {\includegraphics[scale=0.6]{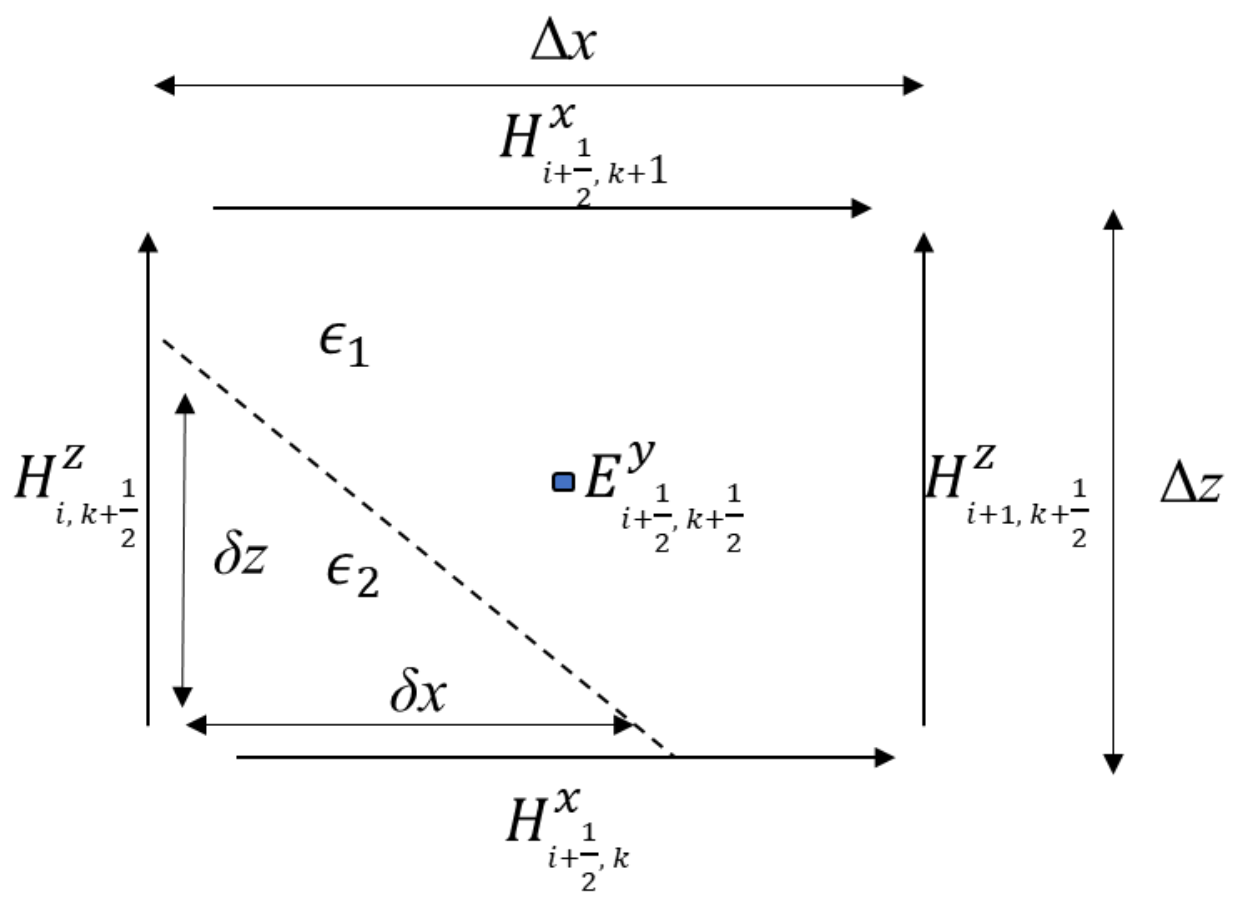}}
\caption{{Geometry for cells found in the interface between the two media.}}
\label {fig:CP}
\end{figure}

On each side of the slab, we place a rough surface with a Gaussian roughness spectrum, generated as discussed in the previous section. To capture the effect of roughness, we use a contour path (CP) FDTD in the main grid \cite{Yu-Mitral}. For cells that are away from the interfaces between the two media, we use the standard  TE uniaxial PML FDTD equations \cite{taflove}. For the cells found on the boundary between the two media, similarly to \cite{Yu-Mitral}, we apply Faraday's law  to deform the contour and assign an effective relative permittivity (see Fig.~\ref{fig:CP}):

 \begin{equation}
    \epsilon_{{\rm eff}}= \epsilon_1 \frac{1-\delta x\delta z }{\Delta x\Delta z} + \epsilon_2 \frac{\delta x\delta z }{\Delta x\Delta z}
    \label{eq:effective}
\end{equation}

To extract the reflection and transmission coefficients of a rough slab, we apply a near-to-far-field (NTFF) transformation. To that end, we record the values of the electric and the magnetic fields over the probing lines on the upper and lower layer, as shown in Fig.~\ref{fig:Oblique_Rough}. Consequently, we calculate the equivalent electric and magnetic surface current densities for the upper and lower probing lines, $\mathbf{J}_s = \mp{H_x} \hat{\mathbf{y}}$ and $ \mathbf{M}_s = \mp E_y \hat{\mathbf{x}}  $, respectively. Let  $\mathbf{r'} = x' \mathbf{\hat{x}}+ z_{1,2} \mathbf{\hat{z}}$ be the position vector of a point on the probing line. Let $\mathbf{r}$ be the  position vector at the far-field. For the 2-D case, the electric field ( $E_{\phi_{R,T}}$ for reflection and transmission, respectively), at a distance $r$ in the far field, is estimated as \cite{schneider2010understanding}: 

\begin{equation}
 \hspace{-5.5 mm}    E_{\phi_{R,T}}= -\sqrt{ \frac{j}{8 \pi k r} } e^{-jkr} (L_\theta - \eta_0N_\phi )
    \label{eq:Efar}
\end{equation}

\begin{equation}
     L_{\theta_{R,T}}= \int_{x1}^{x2} M_x \cos\theta \cos\phi ~ e^{jkr\cos\psi} dx'
    \label{eq:L_theta}
\end{equation}

\begin{equation}
    \hspace{-11 mm} N_{\phi_{R,T}}= \int_{x1}^{x2} J_y \cos\phi  ~ e^{jkr\cos\psi} dx'
    \label{eq:N_phi}
\end{equation}

\noindent where $x_1$ and $x_2$ are the beginning and the end of the probing lines, respectively, and $\psi$ is the angle between $\mathbf{r}$ and  $\mathbf{r'}$. Then, we derive the reflection and transmission coefficients,  $ R(\theta),T(\theta)= E_{\phi_{R,T}}(\theta)/E_i$, where $E_i$ is the amplitude of the incident wave.

A remark on the notation: in our simulations, the plane wave propagates in the positive $x, z$ directions, forming an angle $\theta_i$ with the $z$-axis and the normal to the surface unit vector, as shown in Fig.~\ref{fig:Oblique_flat}. The reflected wave in the specular direction should also form an angle $\theta_i$ with respect to the normal vector, and hence an angle of 180 - $\theta_i$ with the $z$-axis. Therefore, one should observe the electric field from 90 to 270 degrees to capture the diffuse reflection and, correspondingly, from -90 to 90 degrees to capture diffuse transmission. With respect to~(\ref{eq:L_theta}) and~(\ref{eq:N_phi}),  this is equivalent to varying $\theta$ from 90 to 180 for reflection, or from 0 to 90 for transmission. The value of $\phi$ remains constant, equal to 0 or 180, for the right and left quadrant, respectively.


\section{Numerical Results}

 \begin{figure}[b]
\centering
\hspace*{0 cm} 
    {\includegraphics[scale=0.62]{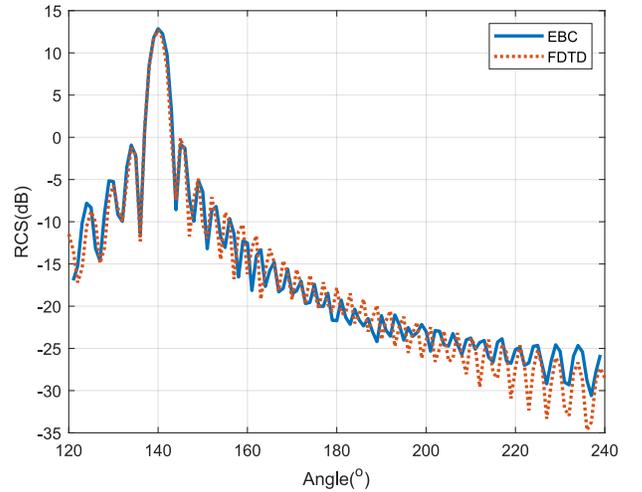}}
\caption{{Comparison between FDTD and FEM for the estimation of the RCS for a one-layer rough surface with $k\sigma_h = 0.1 $.}}
\label {fig:Validation1}
\end{figure}

\subsection{Validation}

Before moving on to the presentation of our results, we validate our model by comparing it to existing methods for the estimation of scattering from rough surfaces. Even though the focus of our research is the impact of roughness in the mm-wave frequency bands used in 5G systems, our approach is directly applicable to other rough surface problems, in areas such as remote sensing.

We evaluate the diffuse scattering from a single rough interface. We compare our results to those derived in \cite{PC_FEM},  where the Finite Element Method (FEM) was employed to estimate the bistatic RCS, $\sigma_{HH} =  \lim_{r\to\infty} 2\pi r |E_{\phi}|^2/|E_i|^2 $. For the validation part only, we generate rough surfaces  with an exponential correlation function \cite{SpanosStochastic}.  

Each surface  has a total length equal to $20\lambda$,  a normalized rms height  $k\sigma_h= 0.1$, and a correlation length equal to $\lambda$. The relative permittivity of the semi-infinite space is $ \epsilon_r = 4- j $. As shown in Fig. ~\ref{fig:Validation1}, there is very good agreement between the two full-wave methods around the specular. Discrepancies, attributed to the grid's discretization or imperfections of the absorber, are observed only for the side lobes. However, the level of the side lobes is considerably smaller than the specular (almost 40 dB below).









\begin{figure}[t!]
\centering
    \subfigure[Reflection coefficient.]
    {\label{fig:PB_R}
    \includegraphics[scale=0.65]{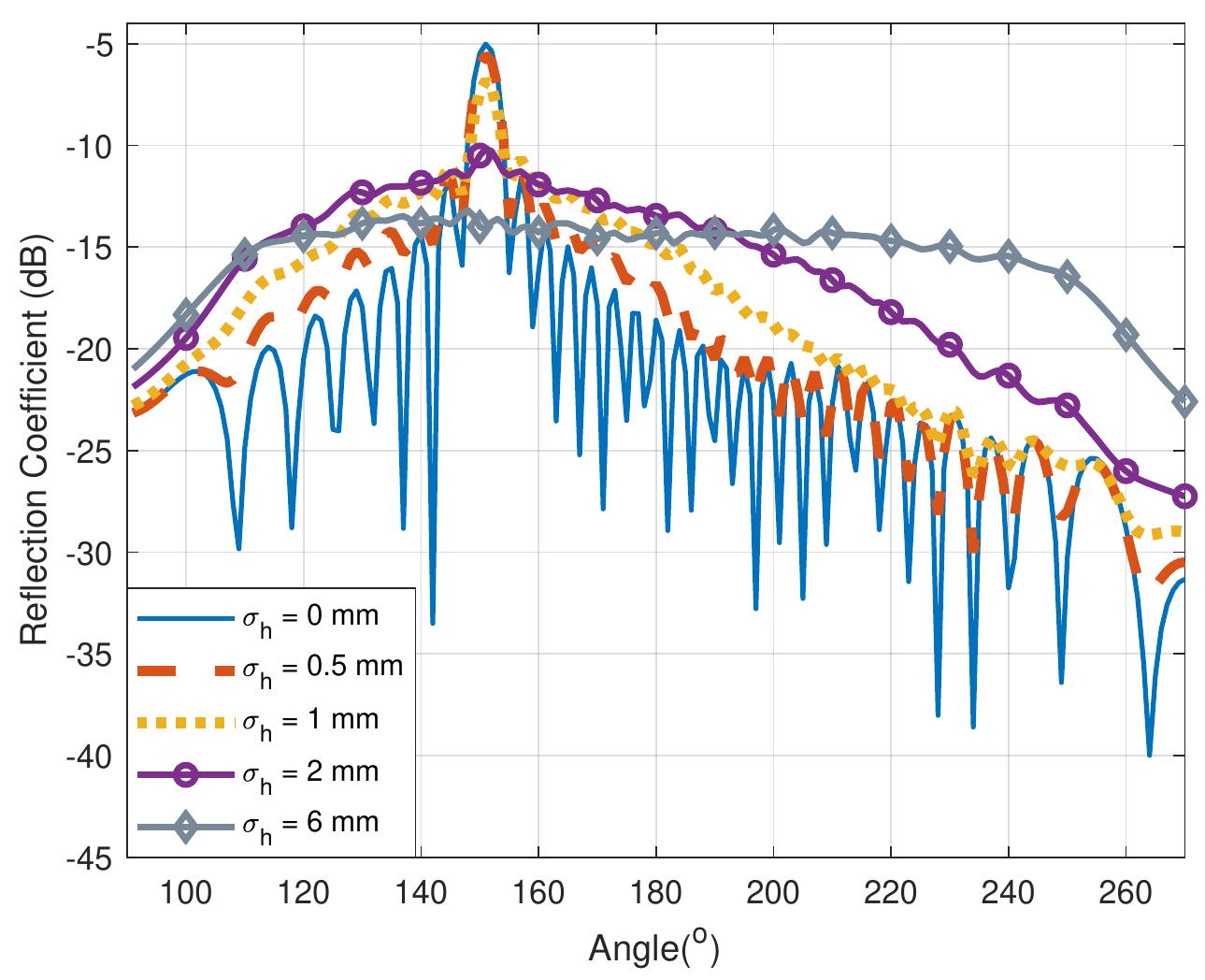}}
    \subfigure[Transmission coefficient.]
    {\label{fig:PB_T}
    \includegraphics[scale=0.66]{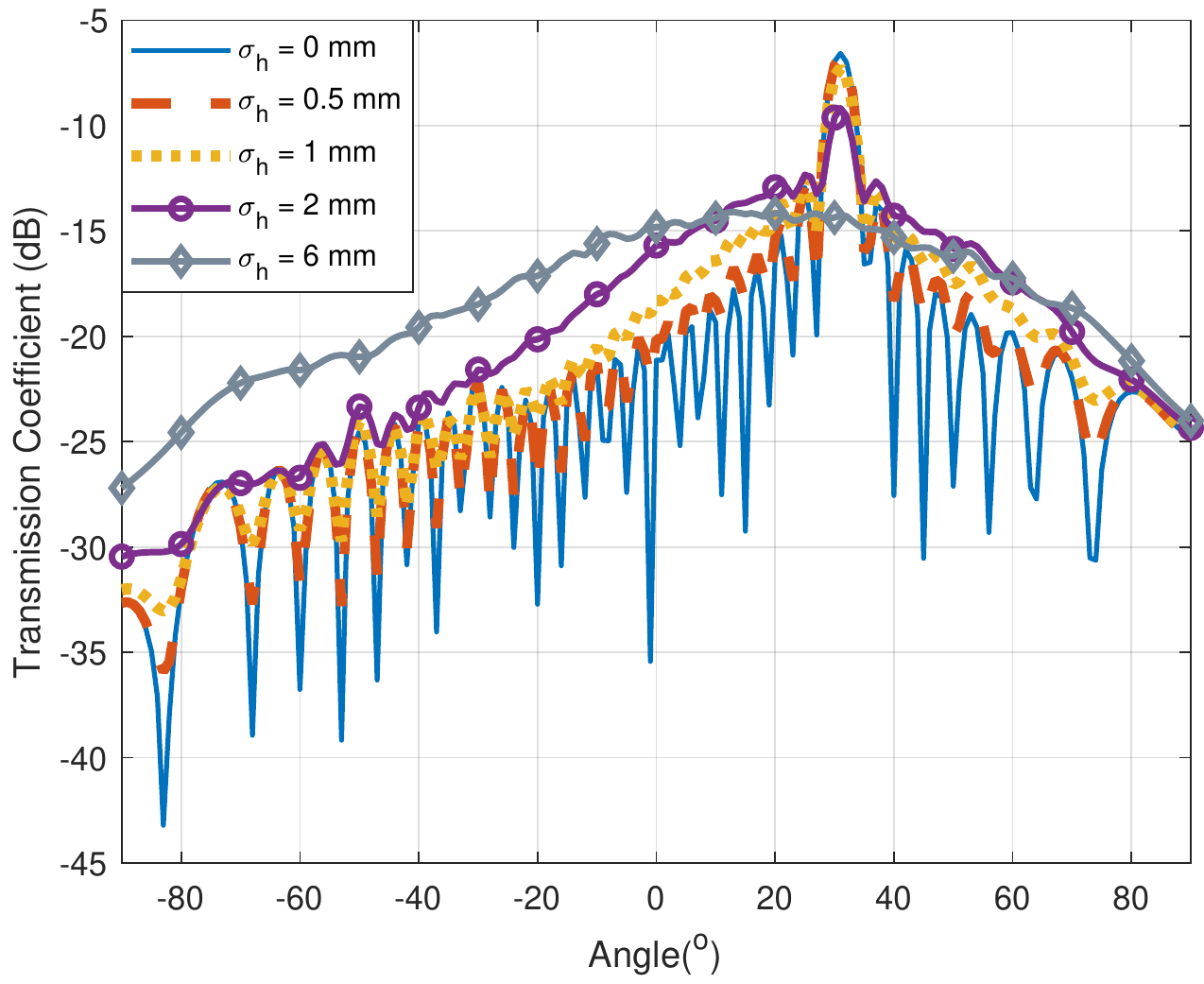}}
\caption{Reflection (a) from and transmission (b) coefficients for a rough plasterboard slab, at 28 GHz, for different values of $\sigma_h$.}
\end{figure}


\subsection{Analysis for construction materials}

We carry out a series of simulations, assuming  TE polarization, for different kinds of materials, different angles of incidence, and different values of $\sigma_h$ for the upper and lower boundaries of a rough slab. For each case, we run 200 Monte Carlo simulations, estimating the average reflection and transmission coefficients from different rough slab realizations. To sample the random variables associated with each FDTD cell, we use the Latin Hypercube sampling method. The resulting rms error, with respect to norm infinity (maximum error for all angles), for the reflection and transmission coefficients is approximately 0.25 dB.

The reflection and the transmission from rough slabs are evaluated for two materials;  wood and  plasterboard. The width of the slabs is 10 cm, for the wooden and the plasterboard slab, respectively. Each material is assumed to be non-magnetic, i.e. $\mu_r=1$, with  a relative permittivity, $ \epsilon_{r_{i}} = \epsilon_i' -j \frac{\sigma_{i} }{\epsilon_0 \omega}$. The conductivity is modeled as $\sigma_i= cf^d$, where $f$ is the frequency of the propagating wave in GHz. The values of $\epsilon_{i}'$, $c$ and $d$, are derived from  the  ITU-R P.2040-1 Recommendation \cite{ITU}, and are illustrated in Table.~\ref{table Materials}. The frequency is set to 28 GHz.

\begin{table}[h]
\centering
\caption{\sc{{Material parameters}}}
\begin{tabular}{ |c||c|c|c|c|} 
 \hline
 \sc{Material} & $\epsilon_i'$ & c & d&  $\sigma  $ (S/m)  \\
 \hline
 \hline
 Wood & 1.99 & 0.0047 & 1.0718  & 0.1672\\
 \hline
 Plasterboard & 2.94 &  0.0116 & 0.7076 & 0.1226 \\
 \hline
\end{tabular}
\label{table Materials}
\end{table}

\begin{figure*}[t]
\centering
\begin{subfigure}[b][ $\theta_i = 30^\circ$]
    {

    \hspace*{-1 mm} 
    \label{fig:Angles_30}
    \includegraphics[scale=0.51]{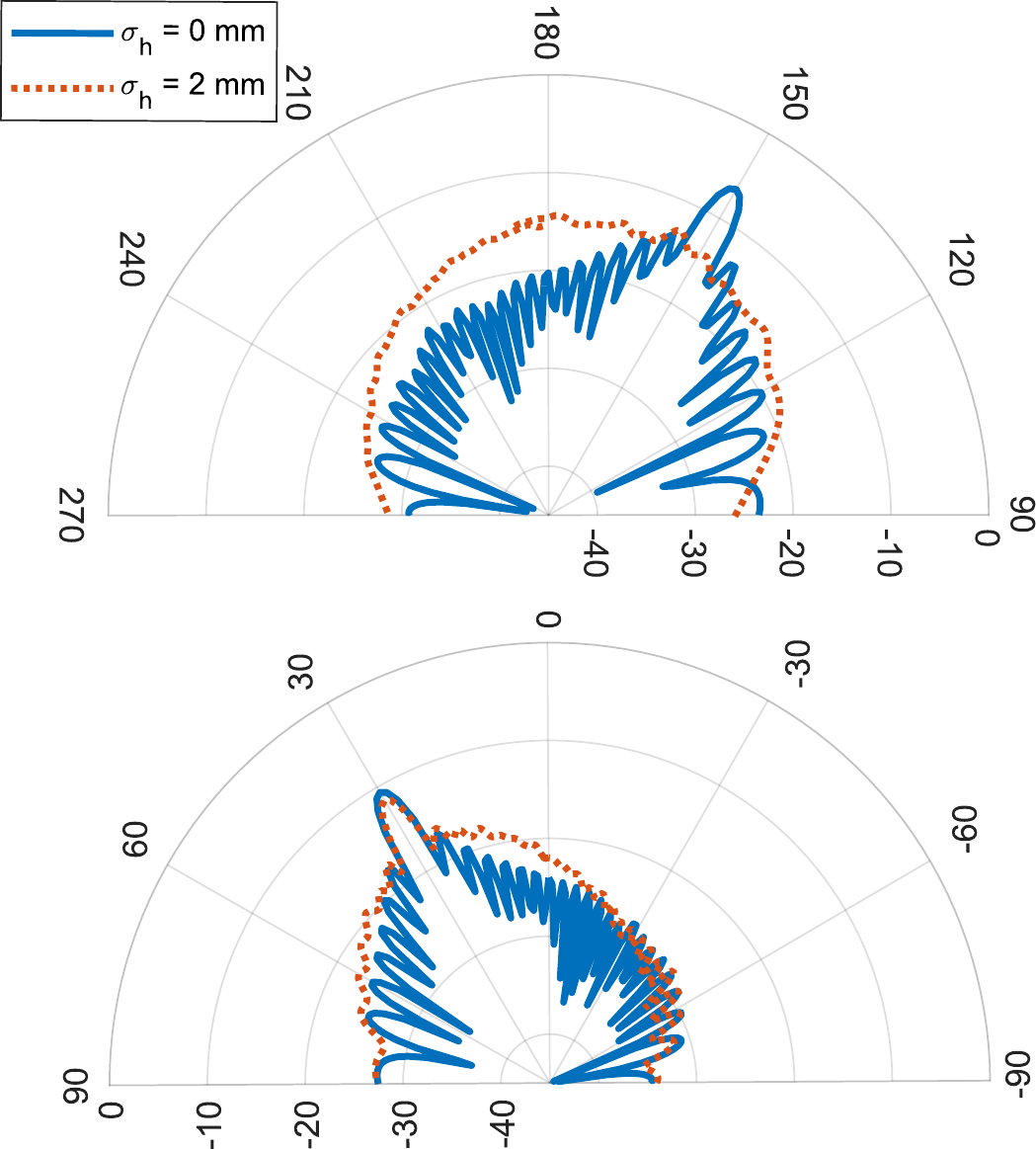}}
    \end{subfigure}
    \hfill
    \begin{subfigure}[b][ $\theta_i = 45^\circ$]
    {\label{fig:Angles_45}
    \includegraphics[scale=0.51]{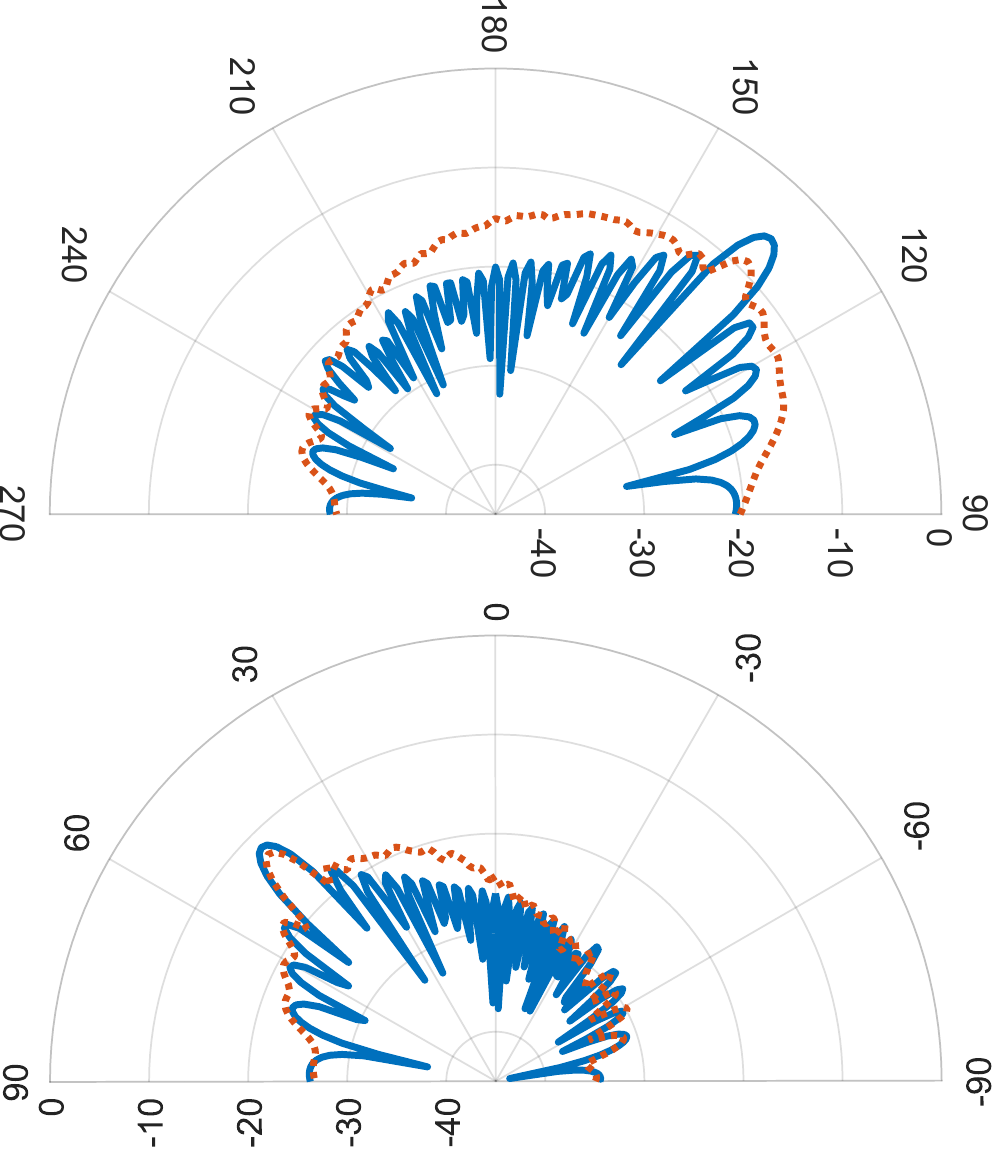}}
    \end{subfigure}
    \hfill
    \begin{subfigure}[b][ $\theta_i = 60^\circ$]
    {\label{fig:Angles_60}
    \includegraphics[scale=0.51]{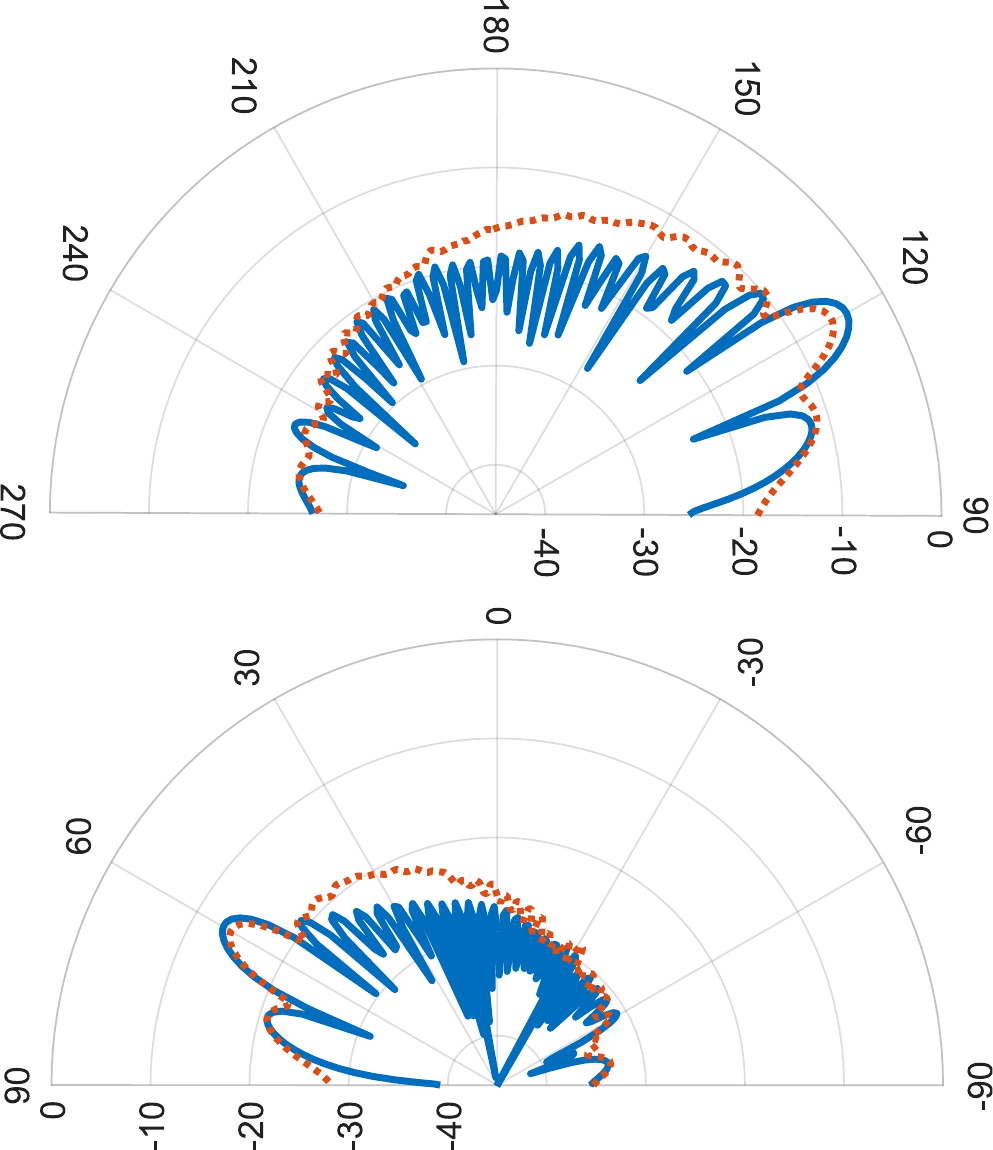}}
    \end{subfigure}
    \hfill
\caption{{Reflection (top) and transmission coefficients (down) for a rough wooden slab, with $\sigma_h = \ 2 \ {\rm mm} $, for $\theta_i = 30^\circ, 45^\circ, 60^\circ$.}}

\end{figure*}

Figures~\ref{fig:PB_R} and ~\ref{fig:PB_T}, show the reflection and transmission coefficients  for a rough plasterboard slab, at an angle of incidence equal to 30 degrees, for four different values of $\sigma_h$. The upper and lower surfaces are assumed to have the same $\sigma_h$. For all cases, the generated surfaces have a correlation length equal to $0.5\lambda$ and a total length equal to 40 correlation lengths ($\Delta x \approx  \lambda/35, N \approx  700$).

For values of $\sigma_h$ below the critical height, the impact of roughness is minor both in reflection and transmission, resulting only in a small attenuation in the specular direction. While the value of $\sigma_h$ increases, a substantial decrease occurs in the specular direction, whereas the scattered field in other directions becomes more intense. For rms heights significantly above the critical height, the specular is vanished.

\begin{table}[h]
\centering
\caption{\sc{{Reduction in Reflection and Transmission Coefficients for $\sigma_h = 2  \ {\rm mm} $. }}}
\begin{tabular}{ |c||c|c|c| } 
 \hline
  \sc Material &$\theta_i = 30^\circ$ & $\theta_i = 45^\circ$ & $\theta_i = 60^\circ$  \\
 \hline
 \hline
 Wood, Reflection  & -4.87 dB & -2.57 dB & -1.64 dB \\
 \hline
 Wood, Transmission & -0.91 dB & -0.97 dB & -1.01 dB \\
 \hline
 Plasterboard, Reflection & -5.15 dB & - 2.99 dB & -2.19 dB \\
 \hline
Plasterboard, Transmission & - 2.57 dB &  -2.6 dB & -2.62 dB \\
 \hline
\end{tabular}
\label{table:Angle}
\end{table}

Figures~\ref{fig:Angles_30} -~\ref{fig:Angles_60}, illustrate the reflection and transmission coefficients for a rough wooden slab, at different angles of incidence, assuming $\sigma_h = 2 \  {\rm mm}$. As the angle of incidence increases, the reduction in the specular decreases, and diffuse scattering is less intense. The opposite holds for the transmission. In Table~\ref{table:Angle}, we illustrate the reduction in dB, for the reflection and transmission coefficients of a  wooden and plasterboard slab, for $\theta_i = 30^{\circ}, 45^{\circ}$ and $60^{\circ}$.

\begin{figure}[t!]
\centering 
    \subfigure[Reflection and  transmission coefficients for a smooth and rough wooden slab.]
    {\label{fig:Wood_30}
    \includegraphics[scale=0.575]{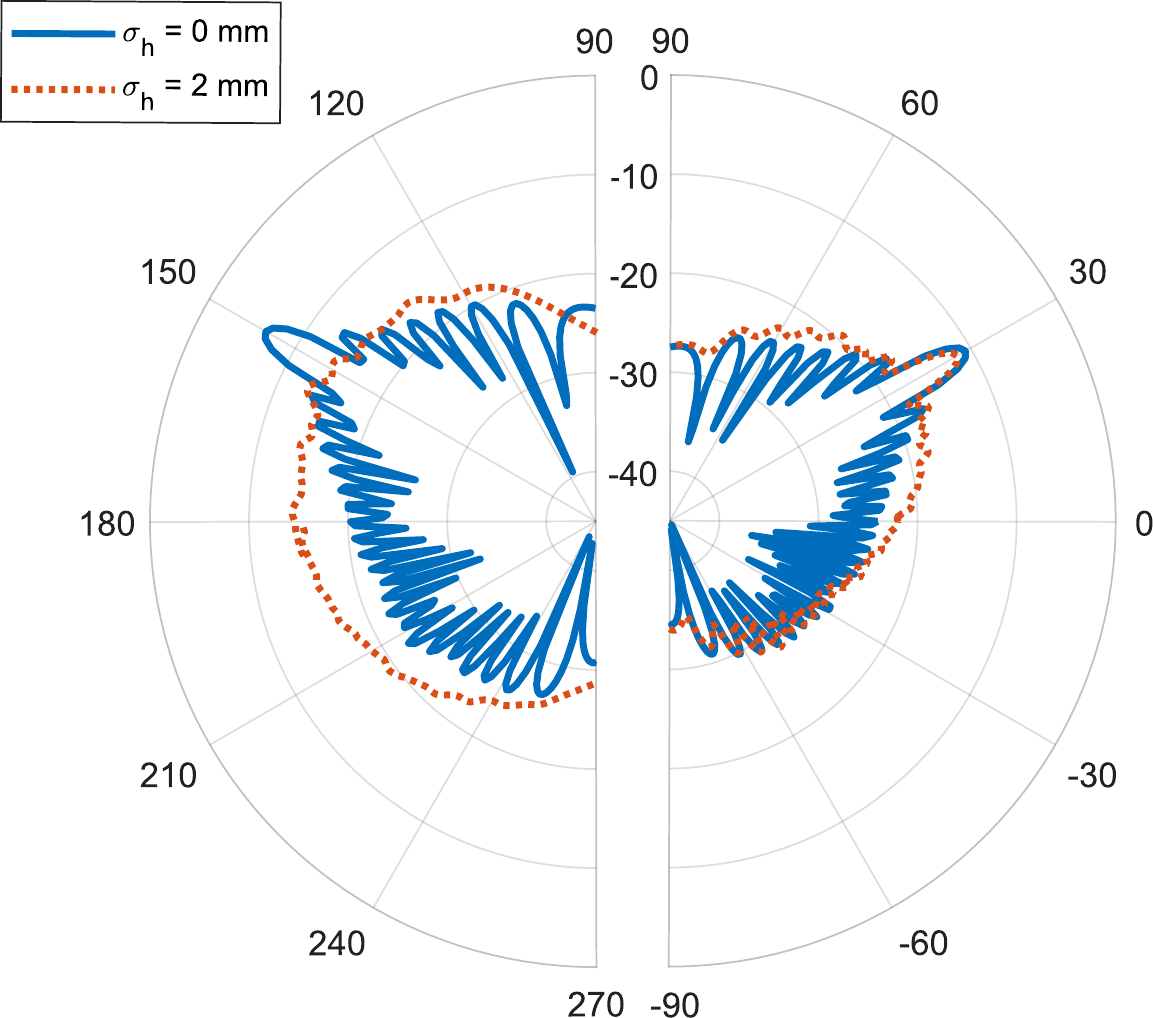}}
    \subfigure[Reflection and transmission coefficients for a smooth and rough plasterboard slab. ]
    {\label{fig:Pb_30}
    \includegraphics[scale=0.575]{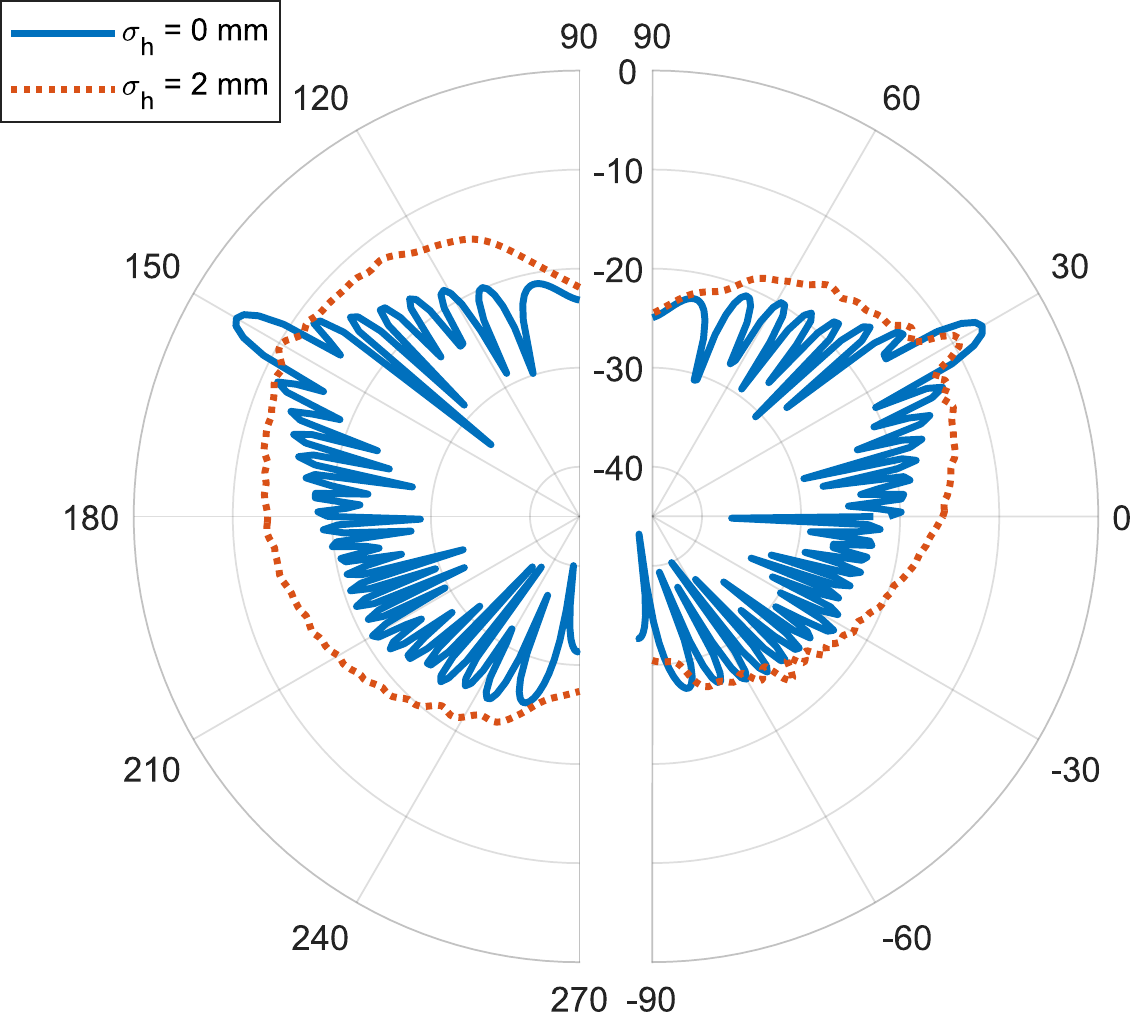}}
\caption{Comparison between the reflection and the transmission coefficients of (a) a wooden and (b) a plasterboard rough slab with $\sigma_h = 2\ {\rm mm}$, at 28  GHz and at $30^\circ$.}
\end{figure}

In Figs.~\ref{fig:Wood_30} and~\ref{fig:Pb_30}, we compare the reflection and transmission coefficients for a wooden and a plasterboard slab, considering an rms height equal to 2 mm. For the first interface, the impact of roughness in scattering remains the same, independently of material parameters. Due to the smaller wavelength in the dielectric,  the influence of roughness is more substantial for the multiply reflected and transmitted components inside the slab. Hence, in reflection we do not observe significant differences for the two materials. However, the reduction in the transmission coefficient, as well as the diffuse transmission in other directions, is more intense for the plasterboard slab. Indeed, from~(\ref{eq:Hc_T}) we find that the critical height in transmission, at $30^\circ$, is approximately 4 and 2 mm for the wood and plasterboard, respectively. Therefore, for the same scale of roughness, the impact of roughness in transmission is more substantial for the plasterboard slab than for the wooden slab.

\begin{figure*}[b!]
\centering

     \hspace*{0.0 cm} 
    \begin{subfigure}[b][Reflected/transmitted fields are attenuated.]
    {\label{fig:R_T_1}
   \includegraphics[scale=0.5]{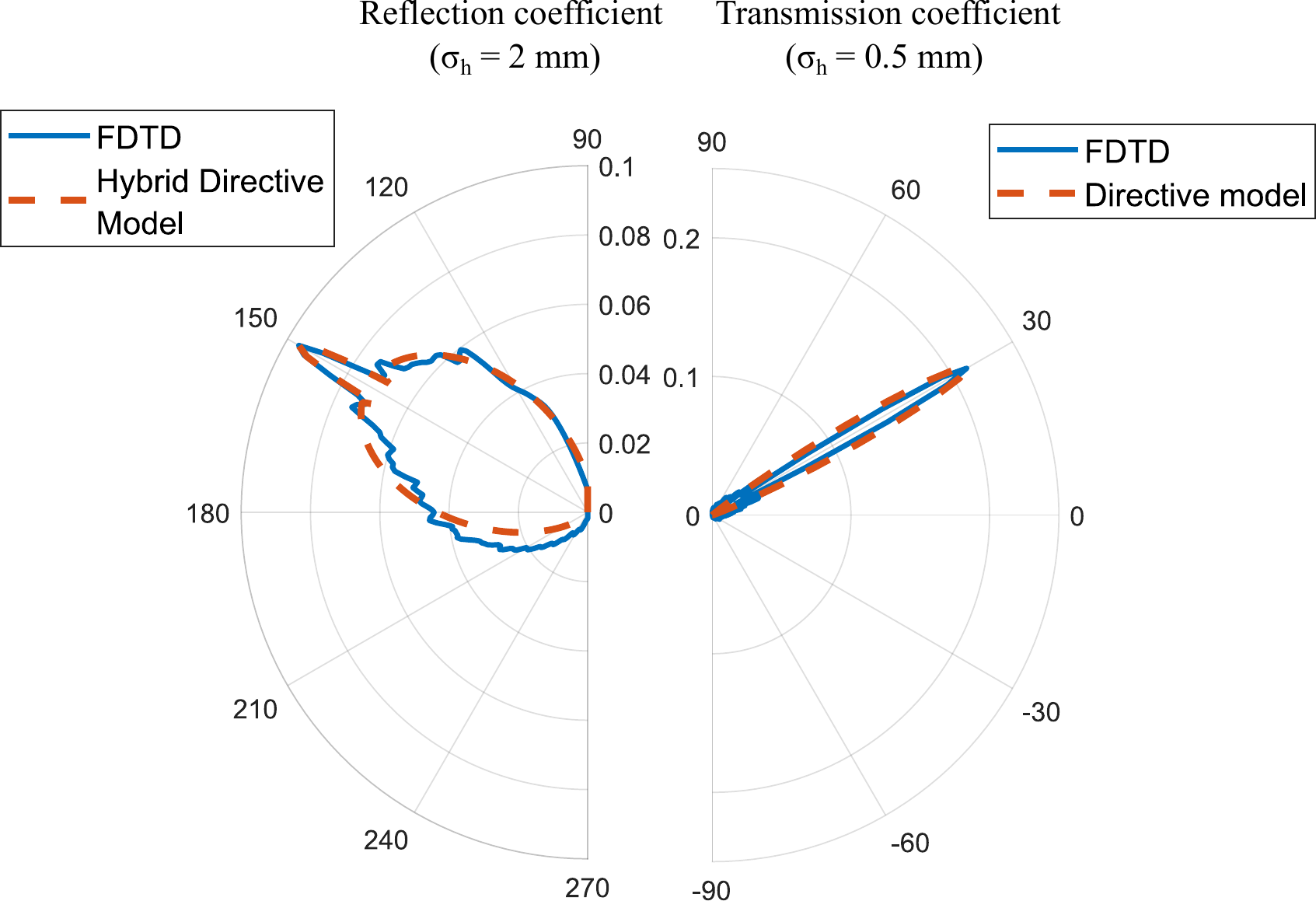}}
    \end{subfigure}
    \hfill
    \begin{subfigure} [b] [Reflected/transmitted fields diffuse to a wide range of angles.]
   {\label{fig:R_T_2}
    \includegraphics[scale=0.5]{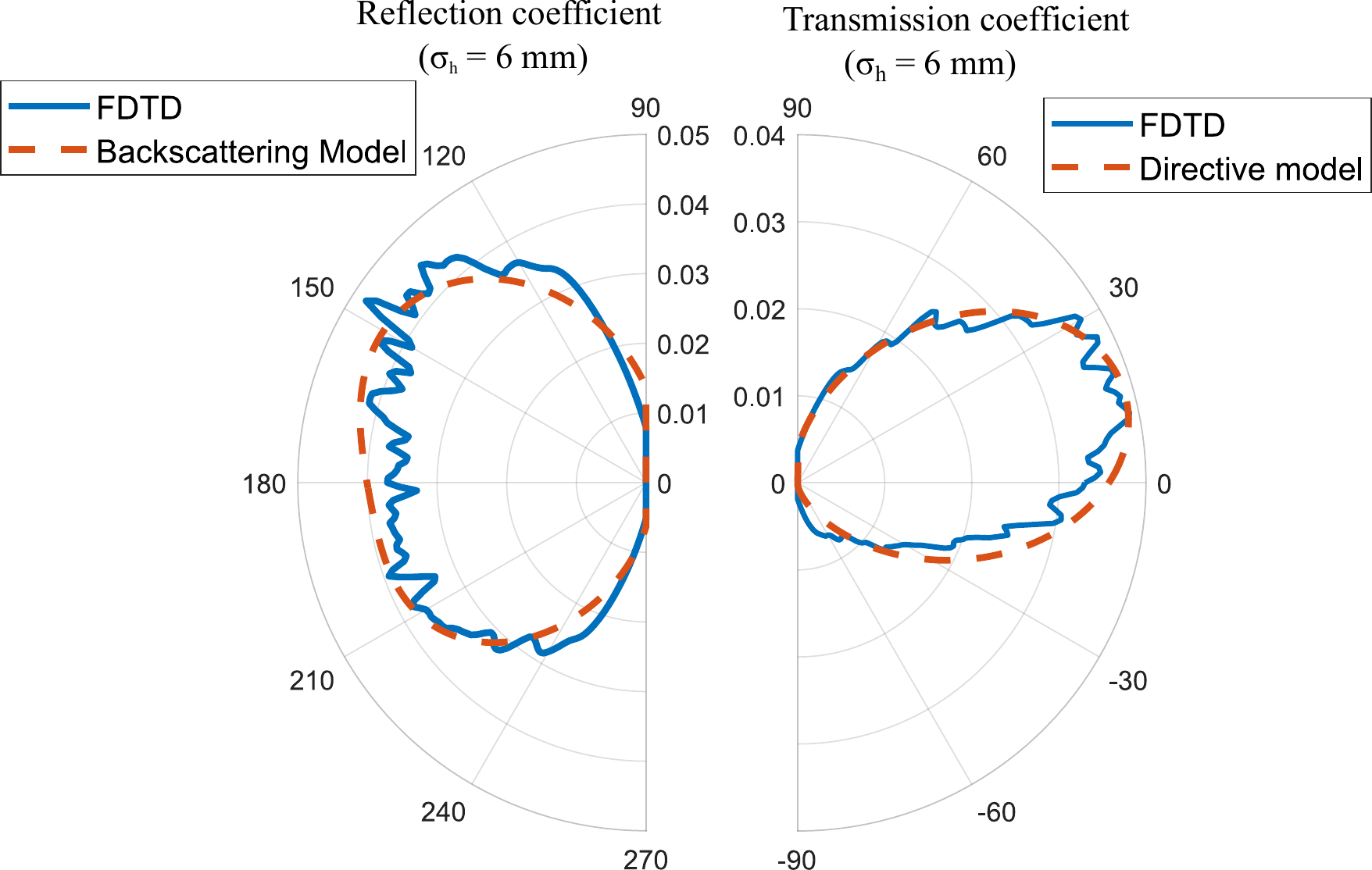}}
     \end{subfigure}
   \hfill
    
\caption{{Approximation of FDTD  reflection and transmission coefficients using the HD, D, BSc and L models.}}
\label{fig: DS_Models}
\end{figure*}

\section{Mapping FDTD Results to Scattering Patterns }

The FDTD results can be embedded in compact models, representing the diffuse scattering properties of a given surface. These can be considered as macromodels of the FDTD simulations, which can be used to efficiently import the FDTD results to other propagation solvers. In \cite{degli2007measurement}, three scattering models were presented; the Lambertian (L), the directive (D), and the backscattering (BSc) model. Let $A(\theta)$ be the radiation pattern of the rough surface in the direction of reflection or transmission. These models are  defined as follows: 

\vspace{-1 mm}
 \begin{equation}
 \hspace{-40.5 mm }    {\rm L}:A(\theta) =  A_o \sqrt{\cos{\theta}}\
     \label{eq:Lambertian}
 \end{equation}

\begin{equation}
 \hspace{-21.5 mm }  {\rm D}:  A(\theta)=  A_{0}\sqrt{ \ \left(\frac{1+\cos{\psi_A}}{2} \right)^{a_A} } \
      \label{eq:Directive} 
 \end{equation}
 
 \begin{multline}
   {\rm BSc}: A(\theta) = \\
   A_{0}\sqrt{ \Lambda\left ( \frac{1+\cos{\psi_A}}{2} \right)^{a_A} + (1-\Lambda)  \left(\frac{1+\cos{\psi_B}}{2} \right)^{a_B}   }
 \label{eq:Backscattering}
\end{multline}
 
\noindent  where  $A_o$ is the amplitude of $A$ and $\psi_{A,B} = \theta - \theta_{A,B}$ ($\theta_B= - \theta_A)$. The parameters $ \theta_{A,B}$ steer the scattering lobes to a certain direction. The parameters $a_A$ and $a_B$ are used to determine the width of the scattering lobes; the greater their value, the narrower the lobe of the reflected/transmitted or of the backscattering beam. 

In scenarios where the specular is preserved and there is also strong diffuse propagation in other directions, we introduce a hybrid directive (HD) model; a directive model with a small lobe approximates the propagation in the specular direction, while one of (\ref{eq:Lambertian})~-~(\ref{eq:Backscattering}) describes the diffuse reflection/transmission. 

We define as $\pmb{x} $, the vector which includes the parameters of $A(\theta)$ to be determined  (e.g. for the directive model, $\pmb{x} = [ A_o, a_A, \theta_A ]$). To determine the appropriate model and compute its parameters, we pass the results from FDTD to a non-linear solver, for the problem:  
\begin{align}
\ \ \ \ \ \ \ \ \ \  \ \!\min_{\pmb{x}}    \ \ \ \   \abs[\Big]{  &A(\theta) - \frac{|E_{R,T}(\theta)|}{|E_i|} }
 \label{eq:Min_Prob}
\end{align}

The minimization problem is solved separately for each model. For instance, for the directive model, the minimization problem has the form:
 
\begin{align}
\!\min_{\pmb{x}=[A_o,a_A, \theta_A ]}       \abs[\Big]{  &A_{0}\sqrt{ \ \left(\frac{1+\cos{(\theta - \theta_A)}}{2} \right)^{a_A} } - \frac{|E_{R,T}(\theta)|}{|E_i|} }
\end{align}
 
We note that the far field patterns, $E_{R,T}$, are sampled at every one degree. Hence, the minimization problem (\ref{eq:Min_Prob}) involves 181 equations. Once the parameters of each model are defined, the model with the minimum mean square error ($MSE$) is selected. The $MSE$ is defined as: 
 \begin{equation}
    \hspace{10mm} MSE =  \frac{1}{181} \sum_{n=1}^{181}  \left ( A(\theta^{(n)}) - \frac{|E_{R,T}(\theta^{(n)})|}{|E_i|}\right)^2   
      \label{eq:Delatphi_T}
 \end{equation}

\noindent where $\theta^{(n)}$ represents the $n$-th angle at which we sample  $E_{R,T}$.

\begin{figure}[h!]
\centering
  {\includegraphics[scale=0.55]{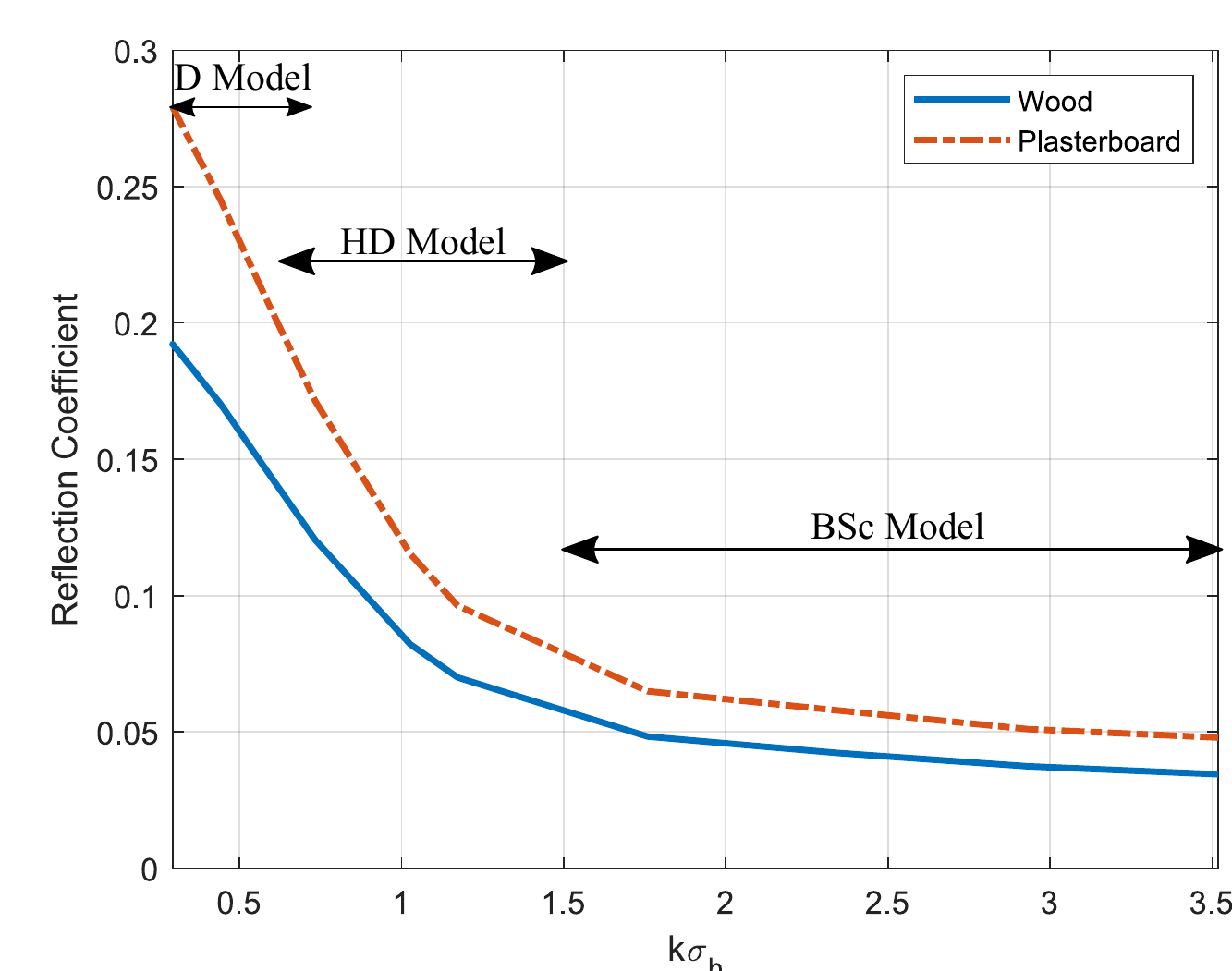}}
\caption{{Reflection coefficient in the specular direction for each slab, for different values of $\sigma_h$, at 28 GHz.}}
\label {fig:Decrease}
\end{figure}

In Fig.~\ref{fig:Decrease}, we show the  absolute value of the specular reflection coefficient for each slab, at 28 GHz and an angle of incidence equal to $30^\circ$. We also indicate the transition between the different scattering models, as the rms height of the two surfaces increases. Contrary to the widely used exponential model that modifies the Fresnel coefficient  \cite{janaswamy2001radiowave}, the reflection coefficient tends to a floor value. For $\sigma_h$ up to approximately $\lambda/10$ (1 mm), roughness results only in an attenuation in the specular direction. Above $\lambda/10$, diffuse reflection becomes significant. Hence, the HD model is used. The specular vanishes after approximately $\lambda/4$ (2.5mm), and a BSc model with two wide lobes is employed to approximate the diffuse reflection.

Tables~\ref{table: Pb_Reflection} and \ref{table: Pb_Transmission} show the models that  approximate the reflection and transmission coefficients of a plasterboard slab, presented in the previous section. For the HD model, we give the parameters for both the models it comprises; (i) the directive, representing the specular propagation, (ii) the model which describes the diffuse scattering. In Fig.~\ref{fig: DS_Models}, we illustrate how these models fit the coefficients obtained from FDTD simulations, distinguishing cases where:  (a) the reflected/transmitted fields propagate primarily in the same directions as for a smooth slab, yet with smaller amplitudes and (b) the reflected/transmitted fields diffuse to a wide range of angles.

\begin{table}[h]
\centering
\caption{\sc{{Reflection Coefficient Models for a Plasterboard Slab.  }}}
\begin{tabular}{ |c||c|c|c|c|c|c|} 
 \hline
 $\sigma_h$ & Model & $A_o$ & $a_A$ & $a_B$ & $\Lambda$ & $\theta_A$ \\
  \hline
  \hline
  0.5 mm & D & 0.2792 & 1000 & - & - & 150 \\
 
 \hline
   1 mm & D & 0.2097 & 1000 & - & - & 150 \\
 \hline
  2 mm & HD         &        &      &   &   &   \\
       & (i) D  & 0.0963 & 1000 & - & - & 150 \\
       & (ii) D  & 0.0749 & 15 & - & - & 150 \\
 \hline
 6 mm & BSc & 0.0556   &    8  & 12   & 0.6   & 150   \\
        \hline
\end{tabular}
\label{table: Pb_Reflection}
\end{table}

\begin{table}[h]
\centering
\caption{\sc{{ Transmission Coefficient Models For a Plasterboard Slab.}}}
\begin{tabular}{ |c||c|c|c|c|c|c|} 
 
 \hline
 $\sigma_h$ & Model & $A_o$ & $a_A$ & $a_B$ & $\Lambda$ & $\theta_A$ \\
 \hline
 \hline
 0.5 mm & D & 0.2118 & 1500 & - & - & 30 \\

 \hline
  1 mm & D & 0.1817 & 1500 & - & - & 30 \\
 \hline
  2 mm & D   &   0.1219     &    1500 & -  & -  &  30 \\
 \hline
 6 mm & D & 0.0415 & 10 & - & - & 15 \\
 \hline
\end{tabular}
\label{table: Pb_Transmission}
\end{table}

\section{Integration With Ray-Tracing}

In this section, we present how the FDTD results are embedded into an RT simulator. To account for roughness, we assume that the illuminated rough surfaces act as secondary radiation sources, whose radiation pattern is computed by FDTD. To that end, the extracted diffuse reflection and transmission coefficients, $R(\theta), T(\theta)$, are directly embedded into a shooting and bouncing ray tracing (SBR) simulator \cite{ling1989shooting}. Instead of rays, our simulator launches  triangular cross-sectional ray tubes from the transmitting point, and employs the imaging method to adjust the ray paths and eliminate spurious rays. 

\begin{figure} [t]
\centering
    
    \subfigure[Specular ray tubes.]
    {\label{fig:Ray_Specular}
    \includegraphics[scale=0.6]{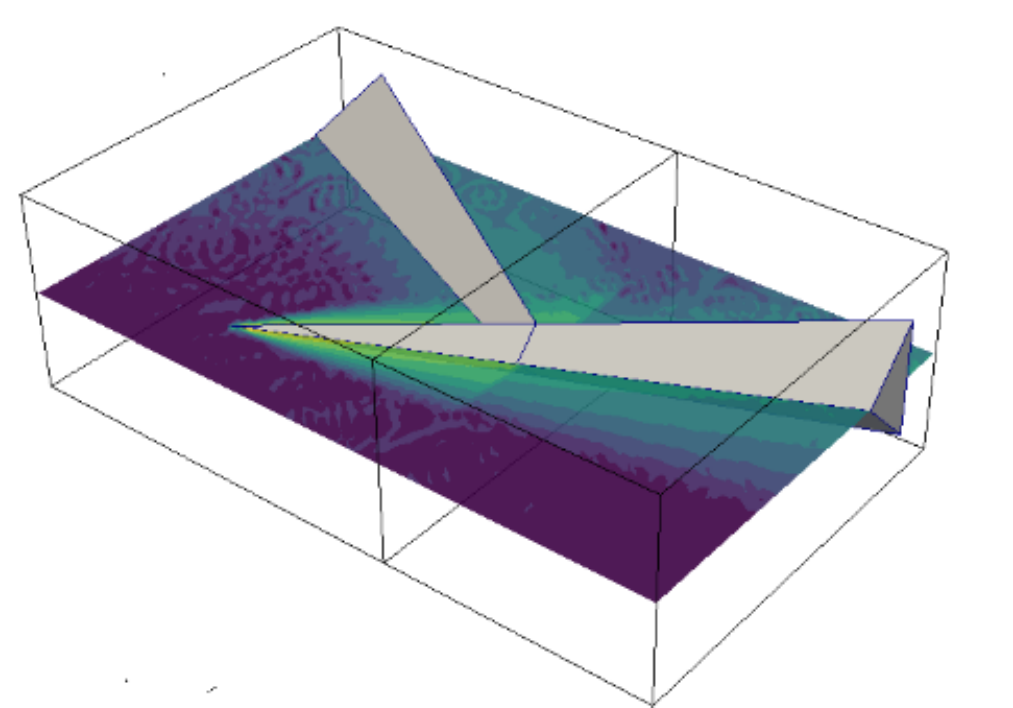}}
    \subfigure[Diffuse transmission ray tubes.]
    {\label{fig:Ray_Diffuse}
    \includegraphics[scale=0.6]{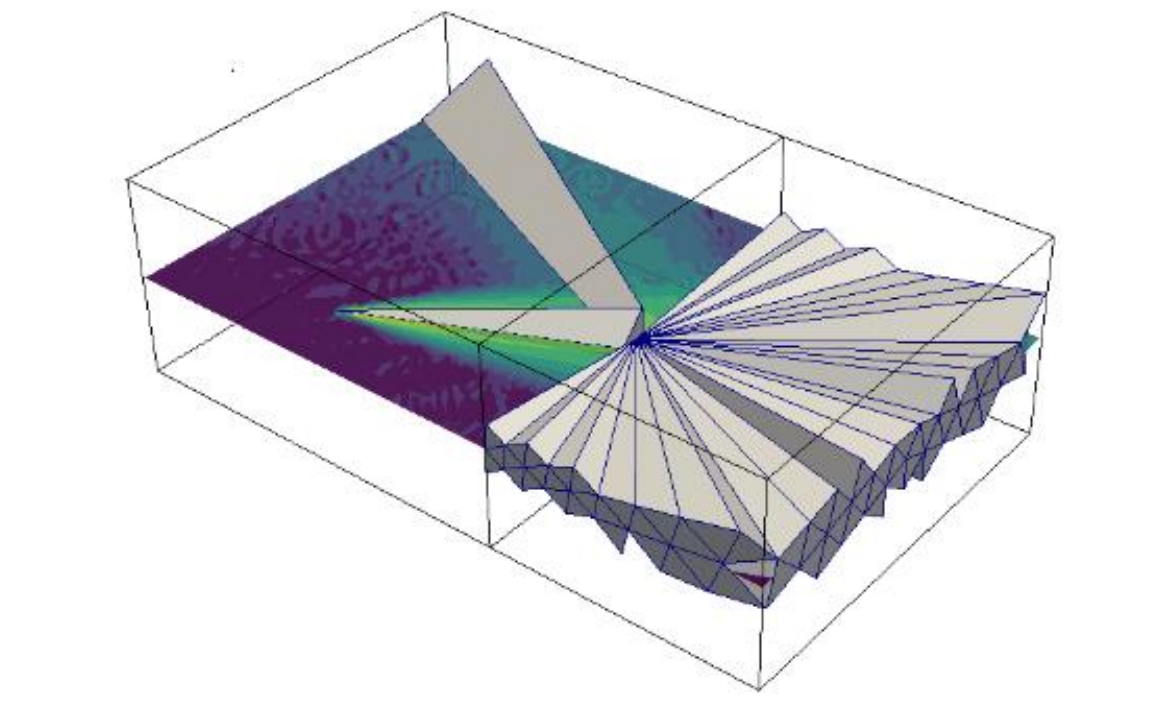}}
 \caption{{Ray tracing with (a) specular ray tubes and (b) diffuse transmission ray tubes.}}
\end{figure}

In the absence of roughness, when a ray tube impinges onto a wall, we consider only the subsequent reflected or transmitted ray tubes in the specular direction, as shown in Fig.~\ref{fig:Ray_Specular}. In the presence of roughness, depending on the scale of roughness, we distinguish the following cases: (i) either the subsequent reflected/transmitted ray tubes  are omitted  or (ii) in cases of pronounced roughness, we generate additional ray tubes according to the reflection/transmission radiation pattern.  Figure ~\ref{fig:Ray_Diffuse} illustrates a case in which the ray tube in transmission is replaced by diffuse ray tubes; similarly one can handle diffuse reflection.
 

\begin{figure}[t]
\centering
    {\includegraphics[scale=0.9]{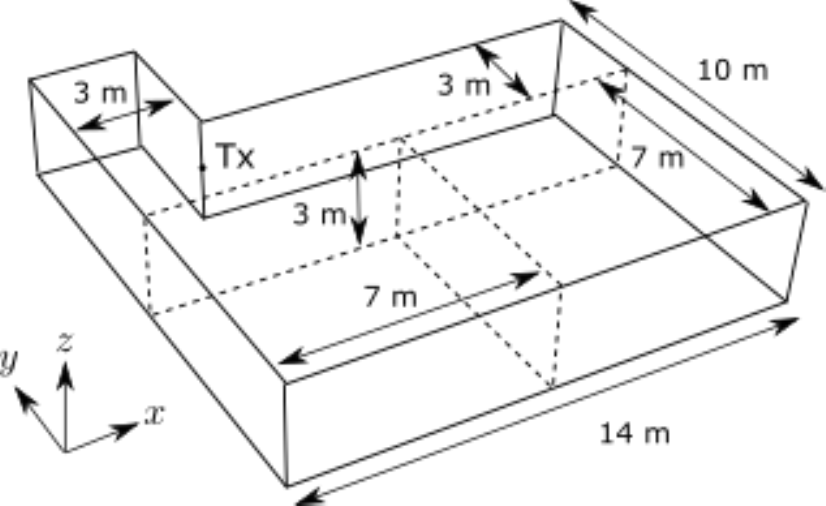}}
\caption{Indoor environment for the ray-tracing simulation.}
\label {fig:Room}
\end{figure}

To demonstrate the influence of diffuse reflection and transmission in mm-wave propagation,  we simulate the received power distribution in an indoor environment. The geometry is  shown in Fig.~\ref{fig:Room}. A big room is divided with a plasterboard wall into two sections. The dimensions of each section are $7 \times 7 \times 3$ m. Next to the rooms, there is an L-shaped corridor, whose width is 3 m. The corridor is separated from the two lower  rooms with a plasterboard wall. 

The plasterboard walls, shown with dashed lines in Fig.~\ref{fig:Room}, are assumed to demonstrate Gaussian roughness and to be 10 cm thick. The correlation length is constant, equal to $ \lambda /2 $. We consider 3 different values of the rms height. The reflection/transmission coefficient patterns, at an angle of incidence equal to $30^\circ$, are illustrated in Figs.~\ref{fig:PB_R} and ~\ref{fig:PB_T}, respectively. The rest of the walls, the floor and the ceiling  are assumed to be smooth,  made of concrete and they are modelled as a semi-infinite space.

The transmitter is fixed on the wall at the corridor, at a 1.5 m height, and uses an antenna with a vertically polarized cosine beam pattern: 
 
  \begin{equation}
     G(\theta)=  2(n+1)\cos^n{\theta}
      \label{eq:Beam pattern}
 \end{equation}

\noindent where we choose $n=100$, in order to generate a narrow beam signal with  Full Width at Half Maximum (FWHM) equal to $14.5^o$. The transmitting power is 20 dBm and the operation frequency is 28 GHz. For each ray tube we allow up to 4  reflections, 2 transmissions and 1 diffuse reflection/transmission. We note that subsequent reflections and transmissions after the diffuse reflection/transmission are also considered in the ray-tracing simulation.

\begin{figure}[h!]
\centering
\begin{tikzpicture}
\centering
  \
  \node at(-0.25,0)
 {\includegraphics[width = 1.0\columnwidth]{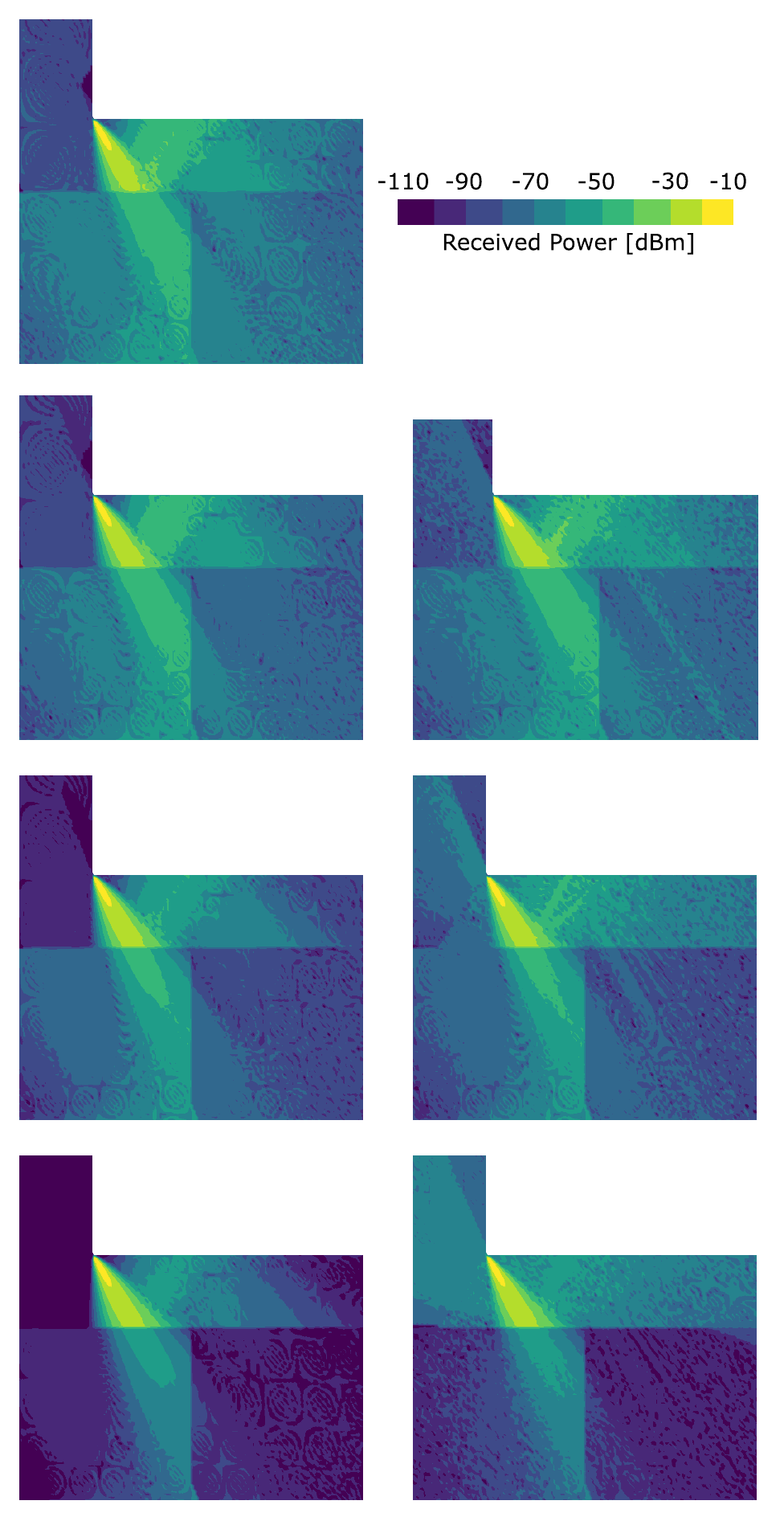}};
\node at (-3.84, 4.30)  {\scalebox{1} {(a)  flat}};

\node at (1.88,0.00)  {\scalebox{1} {(e) $ \sigma_h = 1 \ {\rm mm}$}, with DS};
\node at (-2.68,0.00)  {\scalebox{1} { (b) $ \sigma_h = 1 \ {\rm mm}$, w/o DS}};

\node at (1.88,-4.32)  {\scalebox{1} {(f) $ \sigma_h = 2 \ {\rm mm}$}, with DS};
\node at (-2.68,-4.32)  {\scalebox{1} { (c) $ \sigma_h = 2 \ {\rm mm}$, w/o DS}};
\node at (1.88,-8.72)  {\scalebox{1} {(g) $ \sigma_h = 6 \ {\rm mm}$}, with DS};
\node at (-2.68,-8.72)  {\scalebox{1} { (d) $ \sigma_h = 6 \ {\rm mm}$, w/o DS}};

\end{tikzpicture}
\caption{{Received power distribution, considering (b)-(d) only the attenuation due to roughness, and  (e)-(g) incorporating the FDTD diffuse scattering (DS) and diffuse transmission patterns.}}
\label {fig:Power}
\end{figure}
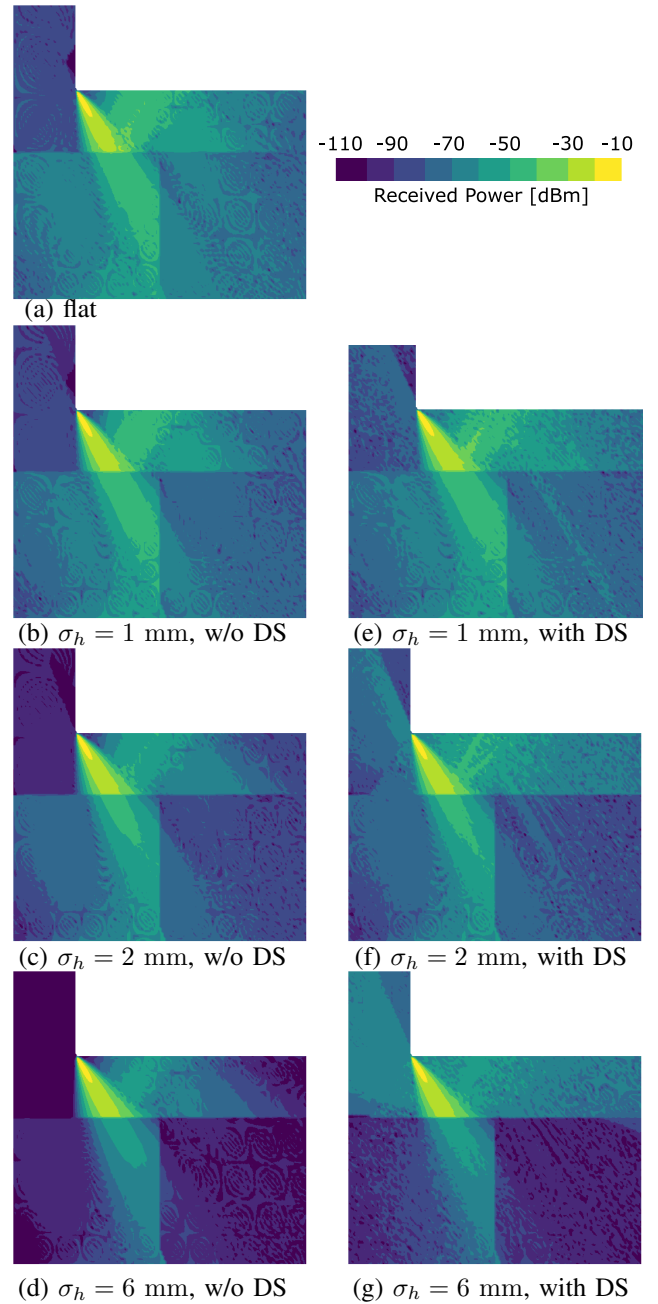

In Fig.~\ref{fig:Power}, we show the received signal strength (RSS) on the horizontal plane, captured by vertically polarized half-wave dipole antennas at a $1.5  \ {\rm m}$ height. In Figs.~\ref{fig:Power}(b) -~\ref{fig:Power}(d), we only consider the attenuation in the specular direction due to roughness, without taking into account any diffuse reflection or diffuse transmission phenomena. In Figs.~\ref{fig:Power} (e) -~\ref{fig:Power}(g), we introduce diffuse ray tubes, according to the diffuse reflection/transmission FDTD coefficients.
\vspace{0mm}

For $\sigma_h$ up to 1 mm, we can observe that generating additional ray-tubes does not have a significant impact on the RSS, as the dominant component of the reflection and transmission patterns are toward the specular direction. However, in cases of pronounced roughness, e.g. $\sigma_h = 2 \ {\rm mm}$ or $\sigma_h= 6  \ {\rm mm}$, the approach of simply attenuating the reflected/transmitted ray tubes is insufficient. The RSS is under-estimated  in the area behind the transmitter, where a considerable part of the corridor is not illuminated. Especially in the 6 mm case, where the non-specular component of the reflection/transmission pattern is significant, there is a noticeable difference. Considering  the diffuse reflection and the subsequent multiple reflections in the corridor, results in improved RSS in the area behind the transmitter, as well as in the upper right corner of the room. 

\section{Conclusion}

In this paper, we illustrated how the FDTD technique can be used to estimate both the reflection from and the transmission through multi-layer rough surfaces. Unlike asymptotic models used for the estimation of diffuse scattering, our approach does not have any constraints regarding the form and degree of roughness. It can be readily applied to extract results for different surface profiles, more complicated geometries or structures with more than two layers. We used our FDTD model to extract reflection and transmission coefficients of rough slabs, rather than those of a single rough boundary, as they are more relevant to indoor propagation scenarios. Furthermore, we showed how the results of the FDTD simulations can be used to derive compact models of the reflection and transmission coefficients of rough slabs. Finally, we integrated our FDTD model with an RT simulator, to illustrate the impact of diffuse scattering and transmission in mm-wave propagation in indoor environments.

\section*{Acknowledgment}
We kindly acknowledge the contribution of  Onassis Foundation in our research.
\vspace{1mm}
\printbibliography

\end{document}